\documentclass[pdftex,11pt,a4paper]{article}

\pdfoutput = 1

\usepackage[T1]{fontenc}
\usepackage{graphicx}
\usepackage{flushend}
\usepackage[numbers,sort&compress]{natbib}
\usepackage[colorlinks,citecolor=blue,urlcolor=blue,linkcolor=blue]{hyperref}
\usepackage{amsfonts}
\usepackage{amsmath}
\usepackage{extarrows}
\usepackage{authblk}

\date{}

\begin{document}

\title{\bf Anomalous transport from holography: Part II}

\author[1]{Yanyan Bu \thanks{yybu@post.bgu.ac.il}}
\author[1,2]{Michael Lublinsky \thanks{lublinm@bgu.ac.il}}
\author[1]{Amir Sharon \thanks{sharon.amir24@gmail.com}}

\affil[1]{\it Department of Physics, Ben-Gurion University of the Negev, \authorcr 
\it Beer-Sheva 84105, Israel}
\affil[2]{\it Physics Department, University of Connecticut, 2152 Hillside Road, \authorcr
\it Storrs, CT 06269-3046, USA}

\maketitle

\begin{abstract}
This is a second study of chiral anomaly induced transport within a holographic model consisting of anomalous $U(1)_V\times U(1)_A$ Maxwell theory in Schwarzschild-$AdS_5$ spacetime. In the first part, chiral magnetic/separation effects  (CME/CSE) are considered in presence of a static spatially-inhomogeneous external magnetic field. Gradient corrections to CME/CSE are analytically evaluated up to third order in the derivative expansion. Some of the third order gradient corrections lead to an anomaly-induced negative $B^2$-correction to the diffusion constant. We also find non-linear in $B$ modifications to the chiral magnetic wave (CMW). In the second part, we focus on the experimentally interesting case of the axial chemical potential being induced dynamically by a constant magnetic and time-dependent electric fields. Constitutive relations for the vector/axial currents are computed employing two different approximations: (a) derivative expansion (up to third order) but fully nonlinear in the external fields, and (b) weak electric field limit but resuming all orders in the derivative expansion. A non-vanishing non-linear axial current (CSE) is found in the first case. Dependence on magnetic field and frequency of linear transport coefficient functions (TCFs) is explored in the second.
\end{abstract}

\newpage


\section{Introduction and summary} \label{intro}

Fluid dynamics \cite{fluid1,fluid2} is an effective long-wavelength description of most classical or quantum many-body systems at nonzero temperature. It is defined in terms of constitutive relations, which relate thermal expectation values of conserved currents to thermodynamical variables and external fields. Derivative expansion in fluid-dynamic variables such as velocity or charge densities accounts for deviations from thermal equilibrium. At each order, the derivative expansion is fixed by thermodynamic considerations and symmetries, up to a finite number of transport coefficients, such as viscosity, diffusion constant  and conductivity. The latter are not calculable from hydrodynamics itself, but have to be determined from underlying microscopic theory or experimentally.

Although fluid dynamics has long history, theoretical foundations of relativistic viscous hydrodynamics are not yet fully established. The Navier-Stokes hydrodynamics leads to violation of causality: the set of fluid dynamical equations makes it possible to propagate signals faster than light. To overcome this problem, simulations of relativistic hydrodynamics are usually based on phenomenological prescriptions of \cite{Muller,Israel,IS1976,IS1979}, which admix viscous effects from second order derivatives, so to make the fluid dynamical equations causal. Refs. \cite{Muller,Israel,IS1976,IS1979} introduced retardation effects for irreversible currents, which, via equations of motion, become additional degrees of freedom. In other words, one needs to include higher order gradient terms in the derivative expansion in order  to obtain a causal formulation. In general, causality is violated if the derivative expansion is truncated at any \emph{fixed} order. It is supposed to be restored when all order gradient terms are included, which we refer to as \emph{all order resummed} hydrodynamics. Resummed hydrodynamics is UV complete in a sense that it has a well-defined large frequency/momenta limit. Yet it is an effective theory of  hydrodynamic variables only\footnote{In fact there are infinitely many such variables (see Ref. \cite{1502.08044} for a discussion).}, which emerges after most of the degrees of freedom of the underlying microscopic theory are integrated out.

The most general parity-even linear in external fields and charge density off-shell constitutive relation for a vector current has the following form
\begin{equation}\label{cr}
J^t=\rho,~~~~~~~~~~~~\vec{J}=-\mathcal{D} \vec{\nabla}\rho+ \sigma_e \vec{E}+ \sigma_m \vec{\nabla} \times \vec{B},
\end{equation}
where $\rho$ is a vector charge density and the diffusion $\mathcal{D}$, electric/magnetic conductivities $\sigma_{e/m}$ are functionals of space-time derivatives. In terms of hydrodynamic expansion, the constitutive relation (\ref{cr}) provides all order resummation of gradients of the fluid-dynamic variables (the charge density $\rho$) and external fields ($\vec{E}$ and $\vec{B}$). In momentum space $\mathcal{D}$ and $\sigma_{e/m}$ are functions of frequency $\omega$ and momentum squared $q^2$ (assuming isotropic medium), which we refer to as transport coefficient functions (TCFs). Via inverse Fourier transform, TCFs appear as memory functions in the constitutive relation \cite{km}.

For a holographic charged plasma dual to $U(1)$ Maxwell theory in Schwarzschild-$AdS_5$  TCFs were studied in depth in \cite{1511.08789}. The derivative resummation in the constitutive relation was implemented via the technique of \cite{1406.7222,1409.3095,1502.08044,1504.01370}, which was originally invented to resum all-order  velocity gradients (linear in the velocity amplitude) in the energy-momentum tensor of a holographic conformal fluid\footnote{One might be concerned that the hydrodynamic derivative expansion forms an asymptotic series with zero radius of convergence \cite{1302.0697}.  However, contrary to our linearised study, this conclusion applies to non-linear hydrodynamics in which the number of terms grows factorially with the number of gradients.  What is more important is that  our approach does not rely on explicit resummation of the gradient series and thus is safe from any convergence related uncertainties.}. It is important to stress that this linearisation procedure is a mathematically well-controlled approximation: the perturbative expansion corresponds to a formal expansion in the amplitudes of fluid-dynamic variables and external fields, without any additional assumptions. In this respect, the implemented approximation is identical to that of the linear response theory based on two-point correlators.

Our technique follows closely the original idea of \cite{0712.2456}, which relates  fluid's constitutive relations for the boundary theory to solving equations of motion in the bulk. However, an important new element of our formalism is that it is not based on current conservation (i.e., ``off-shell'' formalism), which makes it essentially different from the ``on-shell'' formalism of \cite{0712.2456}. Constitutive relations and TCFs can be uniquely determined from dynamical components of the bulk equations only, while the constraint component in the bulk is equivalent to continuity equation on the boundary.

Chiral anomalies emerge and play an important role in relativistic QFTs with massless fermions. The anomaly is reflected in three-point functions of currents associated with global symmetries. When the global $U(1)$ currents are coupled to external electromagnetic fields, the triangle anomaly renders the axial current into non-conserved,
\begin{equation} \label{continuity}
\partial_\mu J^\mu=0,~~~~~~~~~~~~~~~~~~\partial_\mu J^\mu_5=12\kappa \vec{E}\cdot \vec{B},
\end{equation}
where $J^\mu$/$J^\mu_5$ are vector/axial currents, and $\kappa$ is an anomaly coefficient. For $SU(N_c)$ gauge theory with a massless Dirac fermion in fundamental
representation, $\kappa=e N_c/(24\pi^2)$ , and $e$ is an electric charge which below will be set to unit.

Presence of triangle anomalies requires modification of usual constitutive relations for the currents. An example of such modification is the chiral magnetic effect (CME) \cite{hep-ph/0406125,0706.1026,0711.0950,0808.3382, 0911.3715}\footnote{See also \cite{Vilenkin:1980fu,hep-ph/9710234,cond-mat/9803346} for earlier related works.}, that is the induction of an electric current along the applied magnetic field. CME relies on  chiral imbalance, which is usually parameterised by an axial chemical potential. Studies of CME can be found in e.g. \cite{0907.5007,1002.2495,1012.1958,1406.1150,1406.3584,1504.05866,1508.01608} based on perturbation theory, in e.g. \cite{0907.0494,0911.1348,0912.2961,1011.3795,1105.0385,1303.6266} within lattice simulations, and in e.g. \cite{0908.4189,0909.4782,0906.5044,0910.3722,1003.2293,1005.1888,1005.2587,1102.4334,
1103.3773,1112.4227,1207.5309,1305.3949,1407.3282,1410.1306,1603.08770} for strongly coupled regime based on the AdS/CFT correspondence \cite{hep-th/9711200,hep-th/9802109,hep-th/9802150}.

The chiral separation effect (CSE)~\cite{hep-ph/0405216,hep-ph/0505072} is another interesting phenomenon induced by the anomalies. It is reflected in separation of chiral charges along external magnetic field at finite density of vector charges. Chiral charges can  be also separated along  external electric field, when both vector and axial charge densities are nonzero, the so-called chiral electric separation effect (CESE)~\cite{1303.7192,1409.6395}.

In heavy ion collisions, experimentally observable effects induced by the anomalies  were discussed in~\cite{1002.0804,1010.0038,1103.1307,1311.2574,1608.00982}. We refer the reader to~\cite{1211.6245,1210.2186,1312.3348,1509.04073,1511.04050} and references therein for comprehensive reviews on the subject of anomalous transports.

In \cite{1608.08595} we went beyond \cite{1511.08789}
focusing on transport properties induced by the chiral anomaly. The holographic model was modified to be anomalous $U(1)_V\times U(1)_A$ Maxwell theory in Schwarzschild-$AdS_5$. Under various approximations, off-shell constitutive relations were derived for vector/axial currents. In a weak external field approximation, all-order derivatives in the vector/axial currents were resummed into six momenta-dependent TCFs: the diffusion, the electric/magnetic conductivity, and three anomaly-induced TCFs. The latter generalise the chiral magnetic/separation effects. Beyond weak external field approximation, nonlinear transports were also revealed when constant background external fields are present. Particularly, the chiral magnetic effect, including all-order nonlinearity in magnetic field, was proven to be exact when all external fields except for a constant magnetic field are turned off. Nonlinear corrections to the currents' constitutive relations due to electric and axial external fields were computed.

In the present work we continue the study of anomaly-induced transports within the holographic model of \cite{1608.08595}. No axial external fields will be turned on in this work. As in \cite{1511.08789,1608.08595} we work in the probe limit so that the currents and energy-momentum tensor decouple. In the dual gravity, the probe limit ignores backreaction of the gauge dynamics on the geometry. The holographic model under study consists of two Maxwell fields in the Schwarzschild-$AdS_5$ black brane geometry. The chiral anomaly is holographically realised via the gauge Chern-Simons actions for both Maxwell fields. Such a holographic setup can be realised via a top-down brane construction of $D4/D8/\overline{D8}$ \cite{hep-th/0412141}.

Before diving into the details presented in the following sections, we summarise our main results. The paper is split into two largely independent parts. In the first part, we consider a setup in which a static but spatially-varying magnetic field is the only external field that is turned on. Then the constitutive relations for the vector/axial currents are
\begin{equation} \label{jmu varying B1}
J^t=\rho,~~~~~~~~~~~J^i=-\frac{1}{2}\partial_i \rho +\underline{12\kappa \mu_{_5} B_i}- G_i(x=\infty),
\end{equation}
\begin{equation} \label{jmu5 varying B1}
J^t_5=\rho_{_5},~~~~~~~~~~J^i_5=-\frac{1}{2}\partial_i \rho_{_5} +\underline{12\kappa \mu B_i}- H_i(x=\infty),
\end{equation}
where $\rho/\rho_{_5}$ are vector/axial charge densities, the underlined terms in $J^i/J^i_5$ are the chiral magnetic/separation effects. $G_i, H_i$ contain derivatives of $\rho, \rho_{_5}, \vec{B}$  and are defined in section \ref{cme}. It is important to stress that in contrast to the above discussion of linearized hydro,
(\ref{jmu varying B1},\ref{jmu5 varying B1}) are exact, without any approximations for $\rho, \rho_{_5}, \vec{B}$. The nonlinearity of the CME/CSE in external magnetic field $\vec{B}$ is completely accounted for by the chemical potentials $\mu, \mu_{_5}$. The non-derivative part of (\ref{jmu varying B1}) is consistent with the ``non-renormalisability'' of CME \cite{1012.6026,1010.1550,1407.3282,1410.1306}. However, as will be clear from (\ref{Gi cme1},\ref{Hi cme1}), the derivative corrections in $G_i,H_i$ introduce new effects, which do modify the original CME. Particularly, the currents along the direction of $\vec B$ get affected.

When $\rho,\rho_{_5},\vec{B}$ vary slowly from point to point, $G_i,H_i$ can be calculated order-by-order within boundary derivative expansion. Let us introduce a scaling parameter $\lambda$:
\begin{equation} \label{derivative exp}
\partial_\mu=\left(\partial_t,\,\partial_i\right)\longrightarrow \left(\lambda\partial_t,\,\lambda\partial_i\right).
\end{equation}
Then, derivative counting goes  by powers of $\lambda$. Up to second order in derivative expansion, we calculated $G_i, H_i$ and chemical potentials $\mu,\mu_{_5}$ without any further assumptions. Given that these results are rather lengthy, we postpone to present them in section \ref{cme}, see (\ref{Gi cme1},\ref{Hi cme1},\ref{mu/mu5 cme}). At third order $\mathcal{O}(\partial^3)$, for $G_i,H_i$ we calculated only terms that are linear in $\rho,\rho_{_5}$, see (\ref{Gi 3rd},\ref{Hi 3rd}) for a complete listing. Among these third order terms, the diffusion constant $\mathcal{D}_0$ (i.e., the DC limit of the diffusion function $\mathcal{D}$) gets a negative $B$-dependent correction
\begin{equation} \label{anomalous D0}
\mathcal{D}_0=\frac{1}{2}-18(2\log 2-1) \kappa^2B^2.
\end{equation}
To the best of our knowledge, this is the first anomaly-induced correction to the diffusion constant and being negative it happens to violate the universal form of \cite{0806.0110}.

With the third order results for $J^\mu$ and $J^\mu_5$, we also computed the dispersion relation for a free mode that can propagate in the medium:
\begin{equation} \label{dispersion}
\omega=\left[\mp 1+36 \left(1-2\log 2\right)\kappa^2 {\bf B}^2\right]6\kappa\vec{q}\cdot \vec{\bf B}- \left[\frac{1}{2}+18\left(1-2\log 2\right)\kappa^2 {\bf B}^2\right]iq^2- \frac{i}{8}q^4\log 2+\cdots,
\end{equation}
where $\vec{\bf B}$ means a constant magnetic field. The first term in (\ref{dispersion}) represents the chiral magnetic wave (CMW) \cite{1012.6026}. Interestingly, we see nonlinear in ${\bf B}$ corrections to both the speed of CMW and its decay rate. Note that we also expect emergence in (\ref{dispersion}) of the following terms $(\vec{q}\cdot \vec{\bf B})^2$, $q^2(\vec{q}\cdot \vec{\bf B})$, $(\vec{q}\cdot \vec{\bf B})^3$, $q^2 (\vec{q}\cdot \vec{\bf B})^2$ and $(\vec{q}\cdot \vec{\bf B})^4$. However, our ability to determine coefficients of these terms is limited by the undertaken approximations.

In the second part of this work, we focus on a special setup which is experimentally accessible in condensed matter systems\footnote{We thank Dmitri Kharzeev for proposing us this study.}. CME emerges from a nonzero axial chemical potential $\mu_{_5}$, which is usually assumed to have some background profile. It is, however, possible to induce $\rho_{_5}$ (and thus $\mu_{_5}$) dynamically through interplay between the electric and magnetic fields, as is clear from the continuity equation (\ref{continuity}). Specifically, we are ready to consider a constant magnetic field $\vec{\bf B}$ and a time-dependent but spatially-homogeneous electric field $\vec{E}(t)$. For simplicity the charge densities $\rho, \rho_{_5}$ will be assumed to be spatially-homogeneous too\footnote{In principle it is not excluded that the charge densities $\rho,\rho_{_5}$ could be spatially-inhomogeneous. Yet such spatial inhomogeneity would render the derivative resummation highly complicated.}. From (\ref{continuity}), $\rho$ could be set to zero. The constitutive relations for the vector/axial currents are
\begin{equation} \label{jmu exp}
J^t=0,~~~~~~~~~~~~~J^i=E_i+\partial_tE_i +12\kappa \mu_{_5} {\bf B}_i-12 \kappa \epsilon^{ijk} \mathbb{A}_j(1) E_k + \overline{G}_i(x=\infty),
\end{equation}
\begin{equation} \label{jmu5 exp}
J^t_5=\rho_{_5},~~~~~~~~~~~~~~~J^i_5=12\kappa \mu {\bf B}_i-12\kappa \epsilon^{ijk} \mathbb{V}_j(1)E_k +\overline{H}_i(x=\infty),
\end{equation}
where $\mathbb{V}_j(1)$, $\mathbb{A}_j(1)$, $\overline{G}_i$ and $\overline{H}_i$ depend on $\rho_{_5}$, $\vec{E}$ and $\vec{\bf B}$ nonlinearly and will be computed below. Our study is further split into two parts. In section \ref{formal analysis}, $\mathbb{V}_j(1)$, $\mathbb{A}_j(1)$, $\overline{G}_i$ and $\overline{H}_i$ will be evaluated perturbatively within the gradient expansion (\ref{derivative exp}). These perturbative results can be found in (\ref{jmu exp gradient}-\ref{mu/mu5 exp}). In section \ref{AC conductivities}, we will consider another approximation---linearisation of the constitutive relations in the external electric field.

In the linearised regime, we assume the following scaling for $\rho_{_5}$, $\vec{E}$ and $\vec{\bf B}$
\begin{equation} \label{amplitude exp}
\rho_{_5}\sim \mathcal{O}(\epsilon),~~~~~~~~~ \vec{E}(t)\sim \mathcal{O}(\epsilon),~~~~~~~~~
\vec{\bf B}\sim \mathcal{O}(\epsilon^0),
\end{equation}
which will be referred to as amplitude expansion. To linear order in $\epsilon$, the vector/axial currents are
\begin{equation} \label{current exp}
J^t=0,~~~~~\vec{J}=\sigma_e \vec{E}+\tau_1\,\kappa \rho_{_5} \vec{\bf B}+\tau_2\, \kappa^2 \left(\vec{E} \cdot \vec{\bf B}\right)\vec{\bf B};~~~~~~~~~~ J^t_5=\rho_{_5},~~~~~\vec{J}_5=0,
\end{equation}
where $\sigma_e$ is a $q^2=0$ limit of the electric conductivity introduced in (\ref{cr}), while $\tau_{1,2}$ are new TCFs. As with other TCFs, they are functionals of time derivative operator and become functions of frequency $\omega$ in Fourier space,
\begin{equation}
\sigma_e[\partial_t]\longrightarrow \sigma_e(\omega),~~~~~~~~~~~~~~~ \tau_{1,2}[\partial_t]\longrightarrow \tau_{1,2}(\omega).
\end{equation}
At the linear level (in external fields and hydro variables), the transport coefficient functions in \cite{1511.08789,1608.08595} were proved to be frame independent. Along this line of proof, we expect that $\sigma_e,\tau_{1,2}$ are also independent of hydro frame choice. Imposing the continuity equation (\ref{continuity}), the electric current is put on-shell,
\begin{equation} \label{jmu os}
J^i=\sigma_{ij}E_j,~~~~~\sigma_{ij}=\underbrace{\sigma_e}_{\sigma_{\textrm{T}}} \left(\delta_{ij}-\frac{{\bf B}_i {\bf B}_j} {{\bf B}^2}\right)+ \underbrace{\left[\sigma_e-\left(\frac{12}{i\omega} \tau_1-\tau_2\right) \kappa^2 {\bf B}^2\right]}_{\sigma_{\textrm{L}}} \frac{{\bf B}_i {\bf B}_j} {{\bf B}^2},
\end{equation}
where the transverse conductivity $\sigma_{\textrm{T}}$ is not affected by the magnetic field in contrast to the longitudinal conductivity $\sigma_{\textrm{L}}$ which gets corrected by the magnetic field via the chiral anomaly. In section \ref{AC conductivities} the TCFs $\tau_1,\tau_2$ will be first analytically evaluated in the hydro limit and then numerically for arbitrary frequency.

While there is some overlap between our results and the literature, differences between the present study and those of \cite{1407.8162,1410.6399,1504.06566,1603.02624} must be clarified. Utilising the weak electric field approximation (\ref{amplitude exp}), \cite{1407.8162} analytically evaluated the magnetic field dependence of the longitudinal conductivity $\sigma_{\textrm{L}}$ in DC limit, while \cite{1410.6399} calculated its $\omega$-dependence. Back-reaction effects on $\sigma_{\textrm{L}}$ were considered in \cite{1603.02624}. Ref. \cite{1607.06817} performed similar study focusing on time evolution of the induced vector current, given some specially chosen initial profile for the electric field. All the studies \cite{1407.8162,1410.6399,1504.06566,1603.02624} focused on a weak electric field, in which the axial current vanishes. So, our nonlinear results and particularly the axial charge separation current (\ref{jmu5 exp gradient}) appears as new. As for the linearised setup (\ref{amplitude exp}), \cite{1407.8162,1410.6399,1504.06566,1603.02624} imposed the continuity equation and replaced the axial charge density $\rho_{_5}$ in favour of the external electric and magnetic fields, so the vector current there is on-shell. This is in contrast to our off-shell formalism. As we argued in our previous publications \cite{1406.7222,1409.3095,1502.08044,1504.01370,1511.08789}, only off-shell construction reveals transport properties of the system in full. Particularly, there are three independent TCFs ($\sigma_e$ and $\tau_{1,2}$) in the constitutive relation  (\ref{current exp}), all of which we are able to determine separately, compared to only two independent conductivities in (\ref{jmu os}).

Another difference worth mentioning is that we explicitly trace all the effects in the induced current that arise from the relative angle between $\vec{E}(t)$ and $\vec{\bf B}$ fields. This is in contrast to \cite{1410.6399,1607.06817}, which limited their study to the case of parallel fields only, primarily focusing on the longitudinal electric conductivity $\sigma_{\textrm{L}}$. By varying the relative angle between $\vec{E}(t)$ and $\vec{\bf B}$ fields, one can separate the anomaly induced effects (parametrised by $\tau_1$ and $\tau_2$) from the ones that are not related to the anomaly ($\sigma_e$).

The paper is structured as follows. In section \ref{model} we present the holographic model and outline the strategy of deriving the boundary currents from solutions of the anomalous Maxwell equations in the bulk. Section \ref{cme} presents the first part of our study:  CME/CSE with static but varying in space magnetic field. In section \ref{exp setup}, CME/CSE in the presence of constant magnetic and time-varying electric fields are analysed. This study is further split into two subsections. Exploration of nonlinear phenomena in the induced vector/axial currents is done in \ref{formal analysis}. In section \ref{AC conductivities} we focus on the linearised regime (\ref{amplitude exp}) and calculate the dependence of AC conductivity on magnetic field. The last section \ref{conclusion} presents the conclusions. Two Appendices supplement computations of sections \ref{cme} and \ref{exp setup}.

\section{The holographic model: $U(1)_V\times U(1)_A$} \label{model}

The holographic model is the $U(1)_V\times U(1)_A$ theory in the Schwarzschild-$AdS_5$.
The chiral anomaly of the boundary field theory is modelled via the gauge Chern-Simons terms in the bulk action
\begin{equation}
S=\int d^5x \sqrt{-g}\mathcal{L}+S_{\textrm{c.t.}},
\end{equation}
where
\begin{equation}\label{LPVA}
\begin{split}
\mathcal{L}=&-\frac{1}{4} (F^V)_{MN} (F^V)^{MN}-\frac{1}{4} (F^a)_{MN} (F^a)^{MN} +\frac{\kappa\,\epsilon^{MNPQR}}{2\sqrt{-g}}\\
&\times\left[3 A_M (F^V)_{NP} (F^V)_{QR} + A_M (F^a)_{NP}(F^a)_{QR}\right],
\end{split}
\end{equation}
and the counter-term action $S_{\textrm{c.t.}}$ is
\begin{equation}\label{ct VA}
S_{\textrm{c.t.}}=\frac{1}{4}\log r \int d^4x \sqrt{-\gamma}\left[(F^V)_{\mu\nu} (F^V)^{\mu\nu} +(F^a)_{\mu\nu}(F^a)^{\mu\nu}\right].
\end{equation}
The field strengths $(F^V)_{MN}$ and $(F^a)_{MN}$ are defined as
\begin{equation}
(F^V)_{MN}=\partial_MV_N-\partial_NV_M,~~~~~~~~~~~~~~ (F^a)_{MN}=\partial_MA_N-\partial_NA_M.
\end{equation}
$\epsilon^{MNPQR}$ is the Levi-Civita symbol with the convention $\epsilon^{rtxyz}=+1$, and the Levi-Civita tensor is $\epsilon^{MNPQR}/\sqrt{-g}$. Our choice for (\ref{ct VA}) is based on minimal subtraction, that is the counter-term does not make {\it finite} contribution to the boundary currents.

In the ingoing Eddington-Finkelstein coordinates, the spacetime metric is
\begin{equation}
ds^2=g_{_{MN}}dx^Mdx^N=2dtdr-r^2f(r)dt^2+r^2\delta_{ij}dx^idx^j,
\end{equation}
where $f(r)=1-1/r^4$, so that the Hawking temperature (identified as temperature of the boundary theory) is normalised to $\pi T=1$.
On the constant $r$ hypersurface $\Sigma$, the induced metric $\gamma_{\mu\nu}$ is
\begin{equation}
ds^2|_{\Sigma}=\gamma_{\mu\nu}dx^\mu dx^\nu=-r^2f(r)dt^2+r^2\delta_{ij}dx^idx^j.
\end{equation}

Equations of motion for $V$ and $A$ fields are
\begin{equation}\label{eom VAmu}
\textrm{dynamical~~equations}:~~~\textrm{EV}^\mu=\textrm{EA}^\mu=0,
\end{equation}
\begin{equation}\label{eom VAr}
\textrm{constraint~~equations}:~~~\textrm{EV}^r=\textrm{EA}^r=0,
\end{equation}
where
\begin{equation}\label{EV}
\textrm{EV}^M\equiv \nabla_N(F^V)^{NM}+\frac{3\kappa  \epsilon^{MNPQR}} {\sqrt{-g}} (F^a)_{NP} (F^V)_{QR},
\end{equation}
\begin{equation}\label{EA}
\textrm{EA}^M\equiv \nabla_N(F^a)^{NM} +\frac{3\kappa \epsilon^{MNPQR}} {2\sqrt{-g}} \left[(F^V)_{NP} (F^V)_{QR}+  (F^a)_{NP} (F^a)_{QR}\right].
\end{equation}

The boundary currents are defined as
\begin{equation} \label{current definition}
J^\mu\equiv \lim_{r\to\infty}\frac{\delta S}{\delta V_\mu},~~~~~~~~~~~~~
J^\mu_5\equiv \lim_{r\to\infty}\frac{\delta S}{\delta A_\mu},
\end{equation}
which, in terms of the bulk fields, are
\begin{equation}\label{j bct}
\begin{split}
&J^\mu=\lim_{r\to\infty}\sqrt{-\gamma}\,\left\{(F^V)^{\mu M}n_{_M}+ \frac{6\kappa \epsilon^{M\mu NQR}}{\sqrt{-g}}n_{_M} A_N (F^V)_{QR}- \widetilde{\nabla}_\nu (F^V)^{\nu\mu}\log r \right\},\\
&J_5^\mu=\lim_{r\to\infty}\sqrt{-\gamma}\, \left\{(F^a)^{\mu M}n_{_M}+ \frac{2\kappa \epsilon^{M\mu NQR}}{\sqrt{-g}}n_{_M} A_N (F^a)_{QR}- \widetilde{\nabla}_\nu (F^a)^{\nu\mu}\log r \right\},
\end{split}
\end{equation}
where $n_{_M}$ is the outpointing unit normal vector on the slice $\Sigma$, and $\widetilde{\nabla}$ is compatible with the induced metric $\gamma_{\mu\nu}$.

The currents (\ref{current definition}) are defined independently of the  constraint equations (\ref{eom VAr}). Throughout this work, the radial gauge $V_r=A_r=0$ will be assumed. Consequently, in order to completely determine the boundary currents~(\ref{j bct}) it is sufficient to solve the dynamical equations (\ref{eom VAmu}) for the bulk gauge fields $V_\mu,A_\mu$ only, leaving the constraints aside. The constraint equations~(\ref{eom VAr}) give rise to the continuity equations~(\ref{continuity}). In this way, the currents' constitutive relations to be derived below are off-shell.

It is useful to reexpress the currents~(\ref{j bct}) in terms of the coefficients of
near boundary asymptotic expansion of the bulk gauge fields. Near $r=\infty$,
\begin{equation}\label{asmp cov1}
V_\mu=\mathcal{V}_\mu + \frac{V_\mu^{(1)}}{r}+ \frac{V_\mu^{(2)}}{r^2}- \frac{2V_\mu^{\textrm{L}}}{r^2} \log r+\mathcal{O}\left(\frac{\log r}{r^3}\right),~~~~~~~~
A_\mu=
\frac{A_\mu^{(2)}}{r^2}
+\mathcal{O}\left(\frac{\log r}{r^3}\right),
\end{equation}
where
\begin{equation}\label{asmp cov2}
V_\mu^{(1)}=\mathcal{F}_{t\mu}^V,~~~~~~~~~~~4V_\mu^{\textrm{L}}=\partial^\nu \mathcal{F}_{\mu\nu}^V.
\end{equation}
In (\ref{asmp cov1}) the constant term for $A_\mu$ is set to zero given that axial external fields are turned off in our present study.
The holographic dictionary implies that $\mathcal{V}_\mu$ is a gauge potential of external electromagnetic fields $\vec {E}$ and $\vec{B}$,
\begin{equation}
\begin{split}
E_i=\mathcal{F}_{it}^V=\partial_i\mathcal{V}_t-\partial_t \mathcal{V}_i,~~~~ B_i=\frac{1}{2}\epsilon_{ijk}\mathcal{F}_{jk}^V=\epsilon_{ijk}\partial_{j}\mathcal{V}_k.
\end{split}
\end{equation}
When obtaining (\ref{asmp cov1},\ref{asmp cov2}), only the dynamical equations~(\ref{eom VAmu}) were utilised. The near-boundary data $V_\mu^{(2)}$ and $A_\mu^{(2)}$ have to be determined by completely solving (\ref{eom VAmu}) from the horizon to the boundary. The currents (\ref{j bct}) become
\begin{equation}\label{bdry currents}
\begin{split}
J^{\mu}	=\eta^{\mu\nu}(2V_{\nu}^{(2)}+2V^{\textrm{L}}_{\nu}+\eta^{\sigma t} \partial_{\sigma} \mathcal{F}_{t\nu}^V),~~~~~~~~~
J_{5}^{\mu}= \eta^{\mu\nu}2A_{\nu}^{(2)}.
\end{split}
\end{equation}

The remainder of this section is to outline the strategy for deriving the constitutive relations for $J^\mu$ and $J_5^\mu$. To this end, consider finite vector/axial charge densities exposed to external electromagnetic fields. Holographically, the charge densities and external fields are encoded in asymptotic behaviors of the bulk gauge fields. In the bulk, we will solve the dynamical equations (\ref{eom VAmu}) assuming some charge densities and external fields, but without specifying them explicitly.

Following~\cite{1511.08789} we start with the most general static and homogeneous profiles for the bulk gauge fields which solve the dynamical equations (\ref{eom VAmu}),
\begin{equation}\label{homogeneous solution}
V_\mu=\mathcal{V}_\mu-\frac{\rho}{2r^2}\delta_{\mu t},~~~~~~~~~~~~
A_\mu=-\frac{\rho_{_5}}{2r^2}\delta_{\mu t},
\end{equation}
where $\mathcal{V}_\mu,\rho,\rho_{_5}$ are all constants for the moment. Regularity requirement at $r=1$ fixes one integration constant for each $V_i$ and $A_i$. As explained below (\ref{asmp cov2}), the constant in $A_\mu$ is set to zero. Through (\ref{bdry currents}), the boundary currents are
\begin{equation}
J^t=\rho,~~~J^i=0;~~~~~~~~~~~J_5^t=\rho_{_5},~~~J_5^i=0.
\end{equation}
Hence, $\rho$ and $\rho_{_5}$ are identified as the vector/axial charge densities.

Next, following the idea of fluid/gravity correspondence~\cite{0712.2456}, we promote $\mathcal{V}_\mu,\rho,\rho_{_5}$ into arbitrary functions of the boundary coordinates
\begin{equation}
\begin{split}
\mathcal{V}_\mu\to \mathcal{V}_\mu(x_\alpha),~~~~~~~~~~~\rho \to \rho(x_\alpha), ~~~~~~~~~~~
\rho_{_5}\to \rho_{_5}(x_\alpha).
\end{split}
\end{equation}
Then, (\ref{homogeneous solution}) ceases to be a solution of the dynamical equations (\ref{eom VAmu}).
To have them  satisfied, suitable corrections  in $V_\mu$ and $A_\mu$ have to be introduced:
\begin{equation} \label{corrections}
\begin{split}
V_\mu(r,x_\alpha)=\mathcal{V}_\mu(x_\alpha)-\frac{\rho(x_\alpha)}{2r^2}\delta_{\mu t}+ \mathbb{V}_\mu(r,x_\alpha),~~
A_\mu(r,x_\alpha)=
-\frac{\rho_{_5}(x_\alpha)}{2r^2}\delta_{\mu t} + \mathbb{A}_\mu(r,x_\alpha),
\end{split}
\end{equation}
where $\mathbb{V}_\mu,\mathbb{A}_\mu$ will be determined from solving (\ref{eom VAmu}). Appropriate boundary conditions have to be specified. First, $\mathbb{V}_\mu$ and $\mathbb{A}_\mu$ have to be regular over the whole integration interval of $r\in [1,\infty]$. Second, at the conformal boundary $r=\infty$, we require
\begin{equation}\label{AdS constraint}
\mathbb{V}_\mu\to 0,~~~~~~\mathbb{A}_\mu \to 0~~~~~~~\textrm{as}~~~~~~r\to \infty,
\end{equation}
which amounts to fixing external gauge potentials to be $\mathcal{V}_\mu$ and zero (for the axial fields).
Additional integration constants will be fixed by the Landau frame convention for the currents,
\begin{equation}\label{Landau frame}
J^t=\rho(x_\alpha),~~~~~~~~~~~J^t_5=\rho_{_5}(x_\alpha).
\end{equation}
The Landau frame choice can be identified as a residual gauge fixing for the bulk fields.

The vector/axial chemical potentials are defined as
\begin{equation} \label{def potentials}
\begin{split}
\mu&=V_t(r=\infty)-V_t(r=1)=\frac{1}{2}\rho-\mathbb{V}_t(r=1),\\
\mu_{_5}&=A_t(r=\infty)-A_t(r=1)=\frac{1}{2}\rho_{_5}-\mathbb{A}_t(r=1).
\end{split}
\end{equation}
Generically, $\mu,\mu_{_5}$ are nonlinear functionals of densities and external fields.

In terms of $\mathbb{V}_\mu$ and $\mathbb{A}_\mu$, the dynamical equations (\ref{eom VAmu}) are
\begin{equation}\label{eom Vt}
0=r^3\partial_r^2 \mathbb{V}_t+3r^2 \partial_r \mathbb{V}_t+r\partial_r \partial_k \mathbb{V}_k+ 12\kappa \epsilon^{ijk} \left[\partial_r \mathbb{A}_i\left( \partial_j \mathcal{V}_k + \partial_j \mathbb{V}_k\right)+ \partial_r\mathbb{V}_i \partial_j \mathbb{A}_k\right],
\end{equation}
\begin{equation}\label{eom Vi}
\begin{split}
0&=(r^5-r)\partial_r^2 \mathbb{V}_i+(3r^4+1)\partial_r \mathbb{V}_i+2r^3\partial_r \partial_t \mathbb{V}_i-r^3 \partial_r\partial_i \mathbb{V}_t+ r^2\left(\partial_t \mathbb{V}_i- \partial_i \mathbb{V}_t\right)\\
&+r(\partial^2 \mathbb{V}_i - \partial_i\partial_k\mathbb{V}_k)-\frac{1}{2}\partial_i \rho +r^2\left(\partial_t\mathcal{V}_i-\partial_i\mathcal{V}_t\right)+ r\left(\partial^2 \mathcal{V}_i- \partial_i \partial_k \mathcal{V}_k\right)\\
&+12\kappa r^2\epsilon^{ijk}\left(\frac{1}{r^3}\rho_{_5}\partial_j \mathcal{V}_k +\frac{1}{r^3}\rho_{_5}\partial_j\mathbb{V}_k +\partial_r \mathbb{A}_t \partial_j \mathcal{V}_k+\partial_r \mathbb{A}_t \partial_j \mathbb{V}_k\right)\\
&-12\kappa r^2\epsilon^{ijk} \partial_r\mathbb{A}_j\left[\left(\partial_t \mathcal{V}_k- \partial_k \mathcal{V}_t\right)+\left(\partial_t \mathbb{V}_k- \partial_k \mathbb{V}_t\right)+\frac{1}{2r^2}\partial_k \rho\right]\\
&-12\kappa r^2\epsilon^{ijk}\left\{ \partial_r\mathbb{V}_j \left[\left(\partial_t \mathbb{A}_k- \partial_k \mathbb{A}_t\right)+\frac{1}{2r^2}\partial_k \rho_{_5}\right]-\partial_j \mathbb{A}_k\left(\partial_r \mathbb{V}_t +\frac{1}{r^3}\rho \right)\right\},
\end{split}
\end{equation}
\begin{equation}\label{eom At}
0=r^3\partial_r^2 \mathbb{A}_t+ 3r^2 \partial_r \mathbb{A}_t+r\partial_r \partial_k \mathbb{A}_k+ 12\kappa \epsilon^{ijk}\left[ \partial_r \mathbb{V}_i\left( \partial_j \mathcal{V}_k+ \partial_j \mathbb{V}_k\right)+ \partial_r\mathbb{A}_i  \partial_j \mathbb{A}_k\right],
\end{equation}
\begin{equation}\label{eom Ai}
\begin{split}
0&=(r^5-r)\partial_r^2 \mathbb{A}_i+(3r^4+1)\partial_r \mathbb{A}_i+2r^3\partial_r \partial_t \mathbb{A}_i-r^3 \partial_r\partial_i \mathbb{A}_t+ r^2\left(\partial_t \mathbb{A}_i- \partial_i \mathbb{A}_t\right)\\
&+r(\partial^2 \mathbb{A}_i - \partial_i\partial_k\mathbb{A}_k)-\frac{1}{2}\partial_i \rho_{_5} +12\kappa r^2\epsilon^{ijk}\left(\partial_j \mathcal{V}_k +\partial_j \mathbb{V}_k \right) \left(\partial_r \mathbb{V}_t +\frac{1}{r^3}\rho\right) \\
&-12\kappa r^2\epsilon^{ijk} \partial_r\mathbb{V}_j\left[\left(\partial_t \mathcal{V}_k- \partial_k \mathcal{V}_t\right)+\left(\partial_t \mathbb{V}_k- \partial_k \mathbb{V}_t\right)+\frac{1}{2r^2}\partial_k \rho\right]\\
&-12\kappa r^2\epsilon^{ijk}\left\{ \partial_r\mathbb{A}_j \left[\left(\partial_t \mathbb{A}_k- \partial_k \mathbb{A}_t\right)+\frac{1}{2r^2}\partial_k \rho_{_5}\right]- \partial_j \mathbb{A}_k \left(\partial_r \mathbb{A}_t +\frac{1}{r^3}\rho_{_5} \right) \right\}.
\end{split}
\end{equation}
In the following sections we will present solutions to the dynamical equations (\ref{eom Vt}-\ref{eom Ai}) under two independent setups discussed in section \ref{intro}.

\section{CME/CSE with time-independent inhomogeneous magnetic field } \label{cme}

In this section we consider the case in which the magnetic field is the only external field that is turned on. The magnetic field is assumed to be varying in space, but it should be time-independent to avoid creating an electric field. There is no restriction on charge densities $\rho,\rho_{_5}$. From the general results~(\ref{asmp cov1},\ref{asmp cov2}),
\begin{equation}
\mathbb{V}_t, \mathbb{A}_t\sim \mathcal{O}\left(\frac{\log r}{r^3}\right),~~~
\mathbb{V}_i\sim \mathcal{O}\left(\frac{\log r}{r^2}\right), ~~~
\mathbb{A}_i\sim \mathcal{O}\left(\frac{1}{r^2}\right),~~~\textrm{as}~~r\to\infty.
\end{equation}
In obtaining large $r$ estimates for $\mathbb{V}_t$ and $\mathbb{A}_t$, the frame convention (\ref{Landau frame}) was used to fix the coefficients of $1/r^2$ in near-boundary expansion for $V_t,A_t$ (thus those of $\mathbb{V}_t$ and $\mathbb{A}_t$). The dynamical equations (\ref{eom Vt}-\ref{eom Ai}) get simplified,
\begin{equation}\label{eom Vt cme1}
0=r^3\partial_r^2 \mathbb{V}_t+3r^2 \partial_r \mathbb{V}_t+r\partial_r \partial_k \mathbb{V}_k+ 12\kappa \epsilon^{ijk} \left[\partial_r \mathbb{A}_i\left( \partial_j \mathcal{V}_k + \partial_j \mathbb{V}_k\right)+ \partial_r\mathbb{V}_i \partial_j \mathbb{A}_k\right],
\end{equation}
\begin{equation}\label{eom Vi cme1}
\begin{split}
0&=(r^5-r)\partial_r^2 \mathbb{V}_i+(3r^4+1)\partial_r \mathbb{V}_i+2r^3\partial_r \partial_t \mathbb{V}_i-r^3 \partial_r\partial_i \mathbb{V}_t+ r^2\left(\partial_t \mathbb{V}_i- \partial_i \mathbb{V}_t\right)\\
&+r(\partial^2 \mathbb{V}_i - \partial_i\partial_k\mathbb{V}_k)-\frac{1}{2}\partial_i \rho +r \partial_k \mathcal{F}_{ki}^V +12\kappa r^2\epsilon^{ijk} \partial_r \left(\mathbb{A}_t- \frac{\rho_{_5}}{2r^2} \right) \left(\partial_j \mathcal{V}_k +\partial_j \mathbb{V}_k \right)\\
&-12\kappa r^2\epsilon^{ijk}\left\{ \partial_r\mathbb{V}_j \left[\left(\partial_t \mathbb{A}_k- \partial_k \mathbb{A}_t\right)+\frac{1}{2r^2}\partial_k \rho_{_5}\right]-\partial_j \mathbb{A}_k \partial_r \left(\mathbb{V}_t -\frac{\rho}{2r^2} \right)\right\}\\
&-12\kappa r^2\epsilon^{ijk} \partial_r\mathbb{A}_j\left[\left(\partial_t \mathbb{V}_k- \partial_k \mathbb{V}_t\right)+\frac{1}{2r^2}\partial_k \rho\right],
\end{split}
\end{equation}
\begin{equation}\label{eom At cme1}
0=r^3\partial_r^2 \mathbb{A}_t+ 3r^2 \partial_r \mathbb{A}_t+r\partial_r \partial_k \mathbb{A}_k+ 12\kappa \epsilon^{ijk}\left[ \partial_r \mathbb{V}_i\left( \partial_j \mathcal{V}_k+ \partial_j \mathbb{V}_k\right)+ \partial_r\mathbb{A}_i  \partial_j \mathbb{A}_k\right],
\end{equation}
\begin{equation}\label{eom Ai cme1}
\begin{split}
0&=(r^5-r)\partial_r^2 \mathbb{A}_i+(3r^4+1)\partial_r \mathbb{A}_i+2r^3\partial_r \partial_t \mathbb{A}_i-r^3 \partial_r\partial_i \mathbb{A}_t+ r^2\left(\partial_t \mathbb{A}_i- \partial_i \mathbb{A}_t\right)\\
&+r(\partial^2 \mathbb{A}_i - \partial_i\partial_k\mathbb{A}_k)-\frac{1}{2}\partial_i \rho_{_5} +12\kappa r^2\epsilon^{ijk}\left(\partial_j \mathcal{V}_k +\partial_j \mathbb{V}_k \right) \partial_r\left(\mathbb{V}_t -\frac{\rho}{2r^2}\right) \\
&-12\kappa r^2\epsilon^{ijk}\left\{ \partial_r\mathbb{A}_j \left[\left(\partial_t \mathbb{A}_k- \partial_k \mathbb{A}_t\right)+\frac{1}{2r^2}\partial_k \rho_{_5}\right]- \partial_j \mathbb{A}_k \partial_r\left(\mathbb{A}_t +\frac{\rho_{_5}}{2r^2} \right) \right\}\\
&-12\kappa r^2\epsilon^{ijk} \partial_r\mathbb{V}_j\left[\left(\partial_t \mathbb{V}_k- \partial_k \mathbb{V}_t\right)+\frac{1}{2r^2}\partial_k \rho\right].
\end{split}
\end{equation}

For generic profiles of $\rho$, $\rho_{_5}$ and $\vec{B}(\vec{x})$, nonlinearity makes it difficult to solve (\ref{eom Vt cme1}-\ref{eom Ai cme1}). To explore general structure of vector/axial currents, we rewrite the dynamical equations (\ref{eom Vt cme1}-\ref{eom Ai cme1}) into integral forms. In this way, near-boundary asymptotic expansion for $\mathbb{V}_\mu$ and $\mathbb{A}_\mu$ could be extracted from the integral forms of (\ref{eom Vt cme1}-\ref{eom Ai cme1}). For simplicity, we deposit the details into Appendix \ref{app cme}. Substituting near-boundary behavior (\ref{Vt cme}-\ref{Ai cme}) into (\ref{bdry currents}) produces the results (\ref{jmu varying B1},\ref{jmu5 varying B1}). As mentioned below (\ref{jmu varying B1},\ref{jmu5 varying B1}), $G_i, H_i$ are functionals of $\rho,\rho_{_5},\vec{B}$ and are presented in (\ref{Gi cme},\ref{Hi cme}). The formal analyse establishes the structure of $J_\mu/J^5_\mu$, particularly the ``non-renormalisation'' of CME and its gradient corrections.

We proceed with hydrodynamic gradient expansion for $J_\mu/J^5_\mu$. This requires us to perturbatively solve the dynamical equations (\ref{eom Vt cme1}-\ref{eom Ai cme1}) within the boundary derivative expansion (\ref{derivative exp}),
\begin{equation}
\partial_\mu=\left(\partial_t,~\partial_i\right) \longrightarrow \left(\lambda\partial_t,~\lambda\partial_i\right).
\end{equation}
The corrections $\mathbb{V}_\mu$ and $\mathbb{A}_\mu$ are expandable in $\lambda$,
\begin{equation}
\mathbb{V}_\mu=\sum_{n=1}^{\infty}\lambda^n\mathbb{V}_\mu^{[n]},~~~~~~~~~~~~
\mathbb{A}_\mu=\sum_{n=1}^{\infty}\lambda^n\mathbb{A}_\mu^{[n]}.
\end{equation}
At each order in $\lambda$, $\mathbb{V}_\mu^{[n]}$ and $\mathbb{A}^{[n]}$ form a system of ordinary differential equations in $r$-coordinate, which can be solved via direct integration over $r$. The results for $\mathbb{V}_\mu^{[n]}$ and $\mathbb{A}^{[n]}$ up to $n=2$ can be found in Appendix \ref{app cme}, see (\ref{pert VAt 1st}-\ref{pert Ai 2nd}).

Substituting the first order solutions (\ref{pert VAt 1st}-\ref{pert Ai 1st}) into (\ref{Gi cme},\ref{Hi cme}) generates hydrodynamic expansion for $G_i,H_i$ up to second order in gradient expansion (throughout this work, the electromagnetic fields are thought of as of first order in derivative counting)
\begin{equation}\label{Gi cme1}
\begin{split}
G_i(x=\infty)&=-\frac{\pi}{8}\partial_t\partial_i \rho + \left(\frac{3}{2}\pi+3\log 2 \right) \kappa \partial_t \rho_{_5} B_i +18\left(1-2\log 2\right) \kappa^2 \left(\rho_{_5}^2+\rho^2\right) \\
&\times \epsilon^{ijk} \partial_j B_k + 18 (2-3\log 2) \kappa^2 \epsilon^{ijk} \left(\rho_{_5} \partial_j \rho_{_5} B_k +\rho \partial_j \rho B_k \right) +\mathcal{O}(\partial^3),
\end{split}
\end{equation}
\begin{equation}\label{Hi cme1}
\begin{split}
H_i(x=\infty)&=-\frac{\pi}{8}\partial_t\partial_i \rho_{_5} + \left(\frac{3}{2}\pi+3\log 2 \right) \kappa \partial_t \rho B_i +36\left(1-2\log 2\right) \kappa^2 \rho \rho_{_5} \epsilon^{ijk} \partial_j B_k \\
&+18\left(2-3\log 2\right) \kappa^2 \epsilon^{ijk} \left(\rho_{_5} \partial_j \rho B_k + \rho \partial_j \rho_{_5} B_k\right)+\mathcal{O}(\partial^3).
\end{split}
\end{equation}
Meanwhile, the second order results (\ref{pert Vt 2nd},\ref{pert At 2nd}) give rise to the gradient expansion of chemical potentials (\ref{def potentials})
\begin{align} \label{mu/mu5 cme}
&\mu=\frac{\rho}{2}+\frac{1}{16}\left(\pi-2\log2\right)\partial^2\rho- \frac{3}{4} \left(\pi-2\log 2 \right)\kappa B_k\partial_k \rho_{_5}+18\left(1-2\log 2\right) \kappa^2 \rho B^2+\mathcal{O}(\partial^3),\nonumber\\
&\mu_{_5}=\mu(\rho\leftrightarrow \rho_{_5}).
\end{align}

In principle, the second order results (\ref{pert Vt 2nd}-\ref{pert Ai 2nd}) could be inserted into (\ref{Gi cme},\ref{Hi cme}), producing derivative expansion for $G_i(x=\infty)$ and $H_i(x=\infty)$ up to third order. However, at third order $\mathcal{O}(\partial^3)$, computing $G_i, H_i$ becomes quite involved. So, at third order $\mathcal{O}(\partial^3)$ we decided to track only linear in $\rho,\rho_{_5}$ terms. As a result, we are able to identify the first anomalous correction to the diffusion constant $\mathcal{D}_0$ due to magnetic field. The final expressions are
\begin{equation} \label{Gi 3rd}
\begin{split}
G_i^{[3]}(x=\infty)&=\frac{\pi^2}{48}\partial_t^2\partial_i \rho +\frac{1}{16} \left(\pi-2\log 2\right) \partial^2 \partial_i \rho+12 \#_1 \kappa \partial_t^2\rho_{_5} B_i-\frac{\pi^2}{8} \kappa\\
&\times \left[\partial^2 \left(\rho_{_5} B_i \right)- \partial_i \partial_k\left(\rho_{_5} B_k\right)\right]+ \frac{3}{4} (\pi-2\log 2)\kappa \partial_i \left(B_k \partial_k\rho_{_5}\right)\\
&+\underline{18(1-2\log 2) \kappa^2 B^2\partial_i \rho}+18(1-2\log 2)\kappa^2 \rho \partial_i B^2+\mathcal{O}(\rho^2,\rho_{_5}^2,\rho\rho_{_5}),
\end{split}
\end{equation}
\begin{equation} \label{Hi 3rd}
H_i^{[3]}(x=\infty)=G_i^{[3]}(x=\infty)\left(\rho \leftrightarrow\rho_{_5}\right),
\end{equation}
where  $\#_1$ in (\ref{Gi 3rd}) is given by the integral
\begin{equation}
\#_1\equiv \frac{1}{2}\int_1^\infty dy\left[2y\partial_y b_2(y)+ b_2(y)\right]\approx 0.362,
\end{equation}
where $b_2(r)$ is given in (\ref{b2}). The underlined term in (\ref{Gi 3rd}) is a $\kappa^2 B^2$-correction to the diffusion constant $\mathcal{D}_0$. Given that the lowest order anomalous correction to the diffusion constant is negative, it is interesting to explore this effect further for arbitrary magnitude of the magnetic field, which however goes beyond the scope of the present study.

Our results for $J^\mu$ and $J^\mu_5$ can be used to explore dispersion relations for free modes propagating in the chiral medium. We consider a constant magnetic field only. Let us take a plane wave ansatz for the vector/axial charge densities
\begin{equation}
\rho=\delta\rho \exp\left(-i\omega t+\vec{q}\cdot \vec{x}\right),~~~~~~~~~~~~~~~
\rho_{_5}=\delta\rho_{_5} \exp\left(-i\omega t+\vec{q}\cdot \vec{x}\right).
\end{equation}
Then the continuity equation (\ref{continuity}) becomes
\begin{equation}
a\delta \rho+b \delta\rho_{_5}=0,~~~~~~~~b\delta \rho+a \delta\rho_{_5}=0,
\end{equation}
which has a nontrivial solution when and only when
\begin{equation}\label{dispersion equation}
a^2=b^2\Longrightarrow a=\pm b,
\end{equation}
where
\begin{equation}
\begin{split}
a&=-i\omega+\frac{1}{2}q^2 +9\left(\pi- 2\log 2\right)\kappa^2 (\vec{q}\cdot\vec{\bf B})^2 +216\left(1-2\log 2\right)\kappa^3 {\bf B}^2i\vec{q}\cdot \vec{\bf B}+\frac{\pi}{8}i\omega q^2\\
&-\frac{\pi^2}{48}\omega^2q^2- \frac{1}{16}\left(\pi-2\log 2\right)q^4+18 \left(1-2\log 2 \right)\kappa^2 {\bf B}^2 q^2,
\end{split}
\end{equation}
\begin{equation}
\begin{split}
b&=6\kappa i\vec{q}\cdot \vec{\bf B}-\frac{3}{4} \left(\pi-2\log 2\right)\kappa q^2 \vec{q} \cdot \vec{\bf B}- \left(\frac{3}{2}\pi+ 3\log 2\right)\kappa \omega \vec{q}\cdot \vec{\bf B}+12\#_1 \kappa \omega^2 i\vec{q}\cdot \vec{\bf B}\\
&+\frac{3}{4}\left(\pi- 2\log 2\right) \kappa q^2 \vec{q}\cdot \vec{\bf B}.
\end{split}
\end{equation}
Solving (\ref{dispersion equation}) leads to the ${\bf B}$-corrected dispersion relation, as summarised in (\ref{dispersion}).

\section{CME/CSE with constant magnetic and time-dependent electric fields} \label{exp setup}

Creating systems with chiral imbalance ($\mu_5\ne 0$) experimentally is problematic. In this section we consider a special setup in which the axial chemical potential $\mu_{_5}$ is not imposed externally but rather is induced dynamically through the chiral anomaly. This setup is of particular interest due to intriguing possibility for it to be realised experimentally in chiral condensed matter systems. Consider a constant magnetic field $\vec{\bf B}$ and a time-dependent homogeneous electric field $\vec{E}(t)$. We also assume the charge densities to be spatially-homogeneous as well\footnote{While from the continuity equation (\ref{continuity}) the charge densities can still have a nontrivial spatial-dependence, we found that such spatial inhomogeneity of the charge densities would make the gradient resummation out of control.}. The continuity equation (\ref{continuity}) degenerates to
\begin{equation} \label{continuity1}
\partial_t J^t=0,~~~~~~~~~~~~~\partial_t J^t_5=12\kappa \vec{E}\cdot \vec{\bf B},
\end{equation}
which implies that the vector charge density is constant while the axial charge density has nontrivial time dependence inherited from $\vec E(t)$.
The setup under consideration is
\begin{equation} \label{exp setup1}
\rho=0,~~~~~~~~\rho_{_5}=\rho_{_5}(t),~~~~~~~~~\vec{E}=\vec{E}(t),~~~~~~~~~\vec{\bf B} = \textrm{constant}.
\end{equation}
Under the frame convention (\ref{Landau frame}), the corrections $\mathbb{V}_\mu$ and $\mathbb{A}_\mu$ of (\ref{corrections}) depend on $r$ and $t$ only.
As a result, the dynamical equations (\ref{eom Vt}-\ref{eom Ai}) are reduced to
\begin{equation}\label{eom Vt expf}
0=r^3\partial_r^2 \mathbb{V}_t+3r^2 \partial_r \mathbb{V}_t+ 12\kappa \partial_r \mathbb{A}_k {\bf B}_k,
\end{equation}
\begin{equation}\label{eom Vi expf}
\begin{split}
0&=(r^5-r)\partial_r^2 \mathbb{V}_i+(3r^4+1)\partial_r \mathbb{V}_i+2r^3\partial_r \partial_t \mathbb{V}_i+ r^2\partial_t \mathbb{V}_i -r^2E_i +12\kappa r^2 \partial_r \mathbb{A}_t {\bf B}_i \\
&+ \frac{12}{r}\kappa \rho_{_5} {\bf B}_i-12\kappa r^2\epsilon^{ijk} \partial_r \mathbb{A}_j \left(\partial_t \mathbb{V}_k -E_k \right)-12\kappa r^2\epsilon^{ijk} \partial_r\mathbb{V}_j \partial_t \mathbb{A}_k,
\end{split}
\end{equation}
\begin{equation}\label{eom At expf}
0=r^3\partial_r^2 \mathbb{A}_t+ 3r^2 \partial_r \mathbb{A}_t+ 12\kappa \partial_r \mathbb{V}_k {\bf B}_k,
\end{equation}
\begin{equation}\label{eom Ai expf}
\begin{split}
0&=(r^5-r)\partial_r^2 \mathbb{A}_i+(3r^4+1)\partial_r \mathbb{A}_i+2r^3\partial_r \partial_t \mathbb{A}_i+ r^2\partial_t \mathbb{A}_i +12\kappa r^2\partial_r \mathbb{V}_t {\bf B}_i\\
&-12\kappa r^2\epsilon^{ijk} \partial_r\mathbb{V}_j\left(\partial_t \mathbb{V}_k -E_k\right)-12\kappa r^2\epsilon^{ijk} \partial_r\mathbb{A}_j \partial_t \mathbb{A}_k.
\end{split}
\end{equation}

\subsection{Non-linear phenomena: general analysis and derivative expansion} \label{formal analysis}

The objective of this subsection is to show that beyond linearised limit (\ref{scaling}) the setup (\ref{exp setup1}) also induces a non-vanishing axial current $\vec{J}_5$, which has been omitted in the literature. To this end, as in section \ref{cme}, we first give fully nonlinear analysis for the dynamical equations (\ref{eom Vt expf}-\ref{eom Ai expf}), followed by perturbative calculations for $\mathbb{V}_\mu, \mathbb{A}_\mu$ within the derivative expansion (\ref{derivative exp}). All calculational details are deposited into Appendix \ref{app exp}.

As in section \ref{cme} the formal analysis are based on rewriting the dynamical equations (\ref{eom Vt expf}-\ref{eom Ai expf}) into integral form, from which one could deduce near-boundary asymptotic behaviors for $\mathbb{V}_\mu,\mathbb{A}_\mu$. The results can be found in (\ref{Vt exp}-\ref{Ai exp}). Plugged into (\ref{bdry currents}), the near-boundary behavior for $\mathbb{V}_\mu, \mathbb{A}_\mu$ presented in (\ref{Vt exp}-\ref{Ai exp}) is translated into boundary currents (\ref{jmu exp},\ref{jmu5 exp}). Generically, the quantities $\mathbb{V}_i(1)$, $\mathbb{A}_i(1)$, $\overline{G}_i(x=\infty)$ and $\overline{H}_i(x=\infty)$ in (\ref{jmu exp},\ref{jmu5 exp}) cannot be computed analytically. However, as in section \ref{cme}, the formal analyse determines the generic forms for $J^\mu/J_5^\mu$.

Within the gradient expansion (\ref{derivative exp}), we perturbatively solve the dynamical equations (\ref{eom Vt expf}-\ref{eom Ai expf}). Up to second order $\mathcal{O}(\partial^2)$, $\mathbb{V}_\mu, \mathbb{A}_\mu$ are shown in (\ref{Vt cme 2nd}-\ref{Ai cme 2nd}). The perturbative solutions (\ref{Vt cme 2nd}-\ref{Ai cme 2nd}) can be plugged into (\ref{Gi bar},\ref{Hi bar}) to generate hydrodynamic expansion for $J^\mu/J^\mu_5$:
\begin{equation} \label{jmu exp gradient}
\begin{split}
\vec{J}&=12\kappa \mu_{_5}\vec{\bf B}+\vec{E}-\frac{\log 2}{2} \partial_t \vec{E} -\frac{\pi^2}{24} \partial_t^2 \vec{E} -\left(\frac{3}{2}\pi+3\log 2\right)\kappa \partial_t\rho_{_5} \vec{\bf B}\\
& +9\pi^2 \kappa^3 \rho_{_5} \left(\vec{\bf B}\times \vec{E}\right)\times \vec{E}
+12\#_1\kappa \partial_t^2 \rho_{_5} \vec{\bf B}
+\mathcal{O} \left(\partial^4\right),
\end{split}
\end{equation}
\begin{equation} \label{jmu5 exp gradient}
\begin{split}
\vec{J}_5&=12\kappa \mu \vec{\bf B}-36\log 2\, \kappa^2 \rho_{_5} \vec{\bf B}\times \vec{E}+ \frac{3}{2} \left(\pi^2+3\pi \log 2+ 6 \log^22\right) \kappa^2 \partial_t \rho_{_5} \vec{\bf B}\times \vec{E} \\
&-\frac{3}{8}\left(48 \mathcal{C}+\pi^2-12\pi \log 2\right) \kappa^2 \rho_{_5} \vec{\bf B}\times \partial_t \vec{E}+\mathcal{O} \left(\partial^4\right),
\end{split}
\end{equation}
where $\mathcal{C}$ is a Catalan constant and $\#_1$ is known numerically only
\begin{equation} \label{num1}
\#_1 \approx 0.362.
\end{equation}
Up to second order in derivatives $\mathcal{O}(\partial^2)$, the chemical potentials (\ref{def potentials}) are\footnote{While we suspect that the chemical potential $\mu$ is zero to all orders in the gradient expansion, we have not been able to prove that.}
\begin{equation} \label{mu/mu5 exp}
\mu=0+\mathcal{O} \left(\partial^3\right),~~~
\mu_{_5}=\frac{1}{2}\rho_{_5}+\frac{3}{2}\left(\pi-2\log2\right)\kappa \vec{E}\cdot \vec{\bf B} +18\left(1-2\log2\right)\kappa^2\rho_{_5} {\bf B}^2+\mathcal{O} \left(\partial^3\right).
\end{equation}
Evaluated on shell via (\ref{continuity}), the axial current $J_5^i$ is fully
nonlinear in the amplitude of the electric field $\vec{E}(t)$, as clear from (\ref{jmu5 exp gradient}).

\subsection{Linear in $\vec E$ phenomena} \label{AC conductivities}

In the previous subsection we focused on hydrodynamic regime, in which we were able to identify some non-linear phenomena. Below, we proceed with an alternative approximation, that is the weak electric field approximation (\ref{amplitude exp}):
\begin{equation} \label{scaling}
\rho_{_5}(t)\sim \mathcal{O}(\epsilon),~~\vec{E}(t)\sim \mathcal{O}(\epsilon),~~\vec{\bf B} \sim \mathcal{O}(\epsilon^0).
\end{equation}
The scaling of $\rho_{_5}$ follows from the continuity equation (\ref{continuity1}). Both corrections $\mathbb{V}_\mu$ and $\mathbb{A}_\mu$ are  of order $\mathcal{O}(\epsilon)$ too. The dynamical equations (\ref{eom Vt expf}-\ref{eom Ai expf}) get further simplified
\begin{equation} \label{eom Vt exp}
0=r^3\partial_r^2\mathbb{V}_t+3r^2\partial_r \mathbb{V}_t+12\kappa \partial_r \mathbb{A}_k {\bf B}_k,
\end{equation}
\begin{equation} \label{eom Vi exp}
0=(r^5-r)\partial_r^2 \mathbb{V}_i+(3r^4+1)\partial_r \mathbb{V}_i +2r^3 \partial_r \partial_t \mathbb{V}_i+r^2 \partial_t \mathbb{V}_i - r^2 E_i +12\kappa r^2 \left(\partial_r \mathbb{A}_t+ \frac{\rho_{_5}}{r^3}\right){\bf B}_i,
\end{equation}
\begin{equation} \label{eom At exp}
0=r^3\partial_r^2\mathbb{A}_t+3r^2\partial_r \mathbb{A}_t+12\kappa \partial_r \mathbb{V}_k {\bf B}_k,
\end{equation}
\begin{equation} \label{eom Ai exp}
0=(r^5-r)\partial_r^2 \mathbb{A}_i+(3r^4+1)\partial_r \mathbb{A}_i +2r^3 \partial_r \partial_t \mathbb{A}_i+r^2 \partial_t \mathbb{A}_i  +12\kappa r^2 \partial_r \mathbb{V}_t {\bf B}_i.
\end{equation}
Integrating (\ref{eom Vt exp},\ref{eom At exp}) over $r$ once, we get
\begin{equation} \label{VAt exp integrated}
\partial_r \mathbb{V}_t=- \frac{12\kappa}{r^3} \mathbb{A}_k {\bf B}_k,~~~~~~~~~~
\partial_r \mathbb{A}_t=- \frac{12\kappa}{r^3} \mathbb{V}_k {\bf B}_k,
\end{equation}
where the frame convention (\ref{Landau frame}) was used to fix the integration constant. (\ref{VAt exp integrated}) makes it possible to decouple $\mathbb{V}_i, \mathbb{A}_i$ from $\mathbb{V}_t, \mathbb{A}_t$. Consequently, (\ref{eom Vi exp},\ref{eom Ai exp}) become
\begin{equation} \label{eom Vi exp1}
0=(r^5-r)\partial_r^2 \mathbb{V}_i+(3r^4+1)\partial_r \mathbb{V}_i +2r^3 \partial_r \partial_t \mathbb{V}_i+r^2 \partial_t \mathbb{V}_i - r^2 E_i +\frac{12\kappa}{r} {\bf B}_i \left(\rho_{_5} - 12\kappa \mathbb{V}_k {\bf B}_k\right),
\end{equation}
\begin{equation} \label{eom Ai exp1}
0=(r^5-r)\partial_r^2 \mathbb{A}_i+(3r^4+1)\partial_r \mathbb{A}_i +2r^3 \partial_r \partial_t \mathbb{A}_i+r^2 \partial_t \mathbb{A}_i  -\frac{144}{r}\kappa^2 {\bf B}_i (\mathbb{A}_k {\bf B}_k).
\end{equation}
Homogeneity property of (\ref{eom Ai exp1}), combined with the regularity requirement at $r=1$ and vanishing boundary condition at $r=\infty$ for $\mathbb{A}_i$,  fixes $\mathbb{A}_i=0$ completely. From (\ref{VAt exp integrated}), $\mathbb{V}_t=0$. That is,
\begin{equation} \label{VtAi vanish}
\mathbb{V}_t=\mathbb{A}_i=0.
\end{equation}
Therefore, at order $\mathcal{O}(\epsilon)$, the axial current $\vec{J}_5=0$ as read off from (\ref{bdry currents}). This is in contrast with the nonlinear analysis of section \ref{formal analysis}.

The differential equation (\ref{eom Ai exp1}) is {\em linear} in the correction $\mathbb{V}_i$. So, (\ref{eom Ai exp1}) can be solved via the technique developed
in \cite{1406.7222,1409.3095,1502.08044,1504.01370}. The bulk equations reduce to linear inhomogeneous partial differential equations while the inhomogeneous terms are built from boundary derivatives of the fluid-dynamic variables and external fields.  The equations then can be exactly solved using Green function formalism: the bulk fields are decomposed in terms of all possible basic vector structures constructed from the fluid-dynamic variables and external fields. These decomposition coefficients (components of the inverse Green function) are functions of holographic radial coordinate and functionals of boundary derivative operators. The functional dependence of decomposition coefficients on boundary derivative operation encodes all-order linear derivatives in the constitutive relations. Transformed into momentum space, the bulk equations give rise to ordinary differential equations for those decomposition coefficients, which are RG-like equations in AdS space. Solving the RG-like equations completely determines fluid's constitutive relations and all transport coefficients.
Below we implement these steps.

$\mathbb{V}_i$ is decomposed as\footnote{In the decomposition for $\mathbb{V}_i$, one could have included a term $C_4 \vec{E}\times \vec{\bf B}$. However, the coefficient $C_4$ would satisfy a homogeneous ODE. Under the same arguments leading to  $\mathbb{A}_i=0$, $C_4$ has to be zero too.}
\begin{equation}
\mathbb{V}_i=C_1E_i+C_2\kappa\rho_{_5}{\bf B}_i+C_3 \kappa^2\left(\vec{E}\cdot \vec{\bf B}\right){\bf B}_i,
\end{equation}
where
\begin{equation}
C_i=C_i(r,\partial_t)\to C_i(r,\omega),~~~~~~~~i=1,2,3.
\end{equation}
The decomposition coefficients $C_i$'s satisfy partially decoupled ordinary differential equations (ODEs),
\begin{equation} \label{eom C1}
0=(r^5-r)\partial_r^2C_1+(3r^4+1)\partial_rC_1-2i\omega r^3\partial_rC_1-i\omega r^2 C_1 -r^2,
\end{equation}
\begin{equation} \label{eom C2}
0=(r^5-r)\partial_r^2C_2+(3r^4+1)\partial_rC_2-2i\omega r^3\partial_rC_2-i\omega r^2 C_2 +\frac{12}{r}\left(1-12\kappa^2 {\bf B}^2C_2\right),
\end{equation}
\begin{equation} \label{eom C3}
0=(r^5-r)\partial_r^2C_3+(3r^4+1)\partial_rC_3-2i\omega r^3\partial_rC_3-i\omega r^2 C_3 -\frac{144}{r} \left(C_1+\kappa^2{\bf B}^2 C_3\right).
\end{equation}
While $C_1$ does not feel the effect of magnetic field, $C_{2,3}$ have nontrivial dependence on the magnetic field via $\kappa^2{\bf B}^2$.

Near $r=\infty$, pre-asymptotic expansions of $C_i$'s are
\begin{equation}
C_1\to-\frac{1}{r}+\frac{c_1}{r^2}-\frac{i\omega \log r}{2r^2}+ \mathcal{O} \left(\frac{\log r}{r^3}\right),~~~
C_2 \to \frac{c_2}{r^2}+\mathcal{O}\left(\frac{1}{r^3}\right),~~~
C_3 \to \frac{c_3}{r^2}+\mathcal{O}\left(\frac{1}{r^3}\right),
\end{equation}
where $c_i$'s are boundary data and have to be fixed through full solution of (\ref{eom C1}-\ref{eom C3}) from the  horizon $r=1$
to the conformal boundary $r=\infty$.
From (\ref{bdry currents}), the conductivities of (\ref{current exp}) are determined by the boundary data $c_i$'s,
\begin{equation}
\sigma_e=2c_1-\frac{1}{2}i\omega,~~~~~~\tau_1=2c_2,~~~~~~\tau_2=2c_3.
\end{equation}

The ODE for $C_1$ was solved in \cite{1511.08789}. The conductivity $\sigma_e$, which is computed from $C_1$ was completely determined and explored in \cite{1511.08789}, while only $q=0$ limit enters into our current study (the results are quoted below). We therefore focus on the remaining two conductivities $\tau_1,\tau_2$, both induced by the chiral anomaly. As is obvious from (\ref{eom C1}-\ref{eom C3}),  $\tau_1,\tau_2$ depend on the magnetic field  via $\kappa^2{\bf B}^2$.

Using the continuity equation (\ref{continuity1}), the constitutive relations (\ref{current exp}) are put into a linear response form, from which on-shell current-current correlators can be read off. Since the electric field is the only external perturbation that is turned on, it is possible to compute only a subset of all two-point correlators in the theory,
\begin{equation}
\langle J^i J^j\rangle=\underbrace{i\omega \sigma_e}_{G^{\textrm{T}}} \left(\delta_{ij}-\frac{{\bf B}_i {\bf B}_j} {{\bf B}^2}\right)+\underbrace{\left[i\omega \sigma_e-(12\tau_1- i\omega\tau_2)\kappa^2 {\bf B}^2\right]}_{G^\textrm{L}} \frac{{\bf B}_i {\bf B}_j}{{\bf B}^2},
\end{equation}
\begin{equation}
\langle J^t_5 J^i \rangle=-12\kappa {\bf B}_i,
\end{equation}
\begin{equation}
\langle J^t J^i\rangle=\langle J^i_5 J^j\rangle=0,
\end{equation}
where $\langle J^i J^j\rangle$ is split into transverse ($G^{\textrm{T}}$) and longitudinal ($G^{\textrm{L}}$) components with respect to the direction of $\vec{\bf B}$. To determine the remaining current-current correlators we would have to introduce additional field perturbations, particularly an axial external field, which is beyond the scope of this paper.

\begin{figure}[htbp]
\centering
\includegraphics[width=0.496\textwidth]{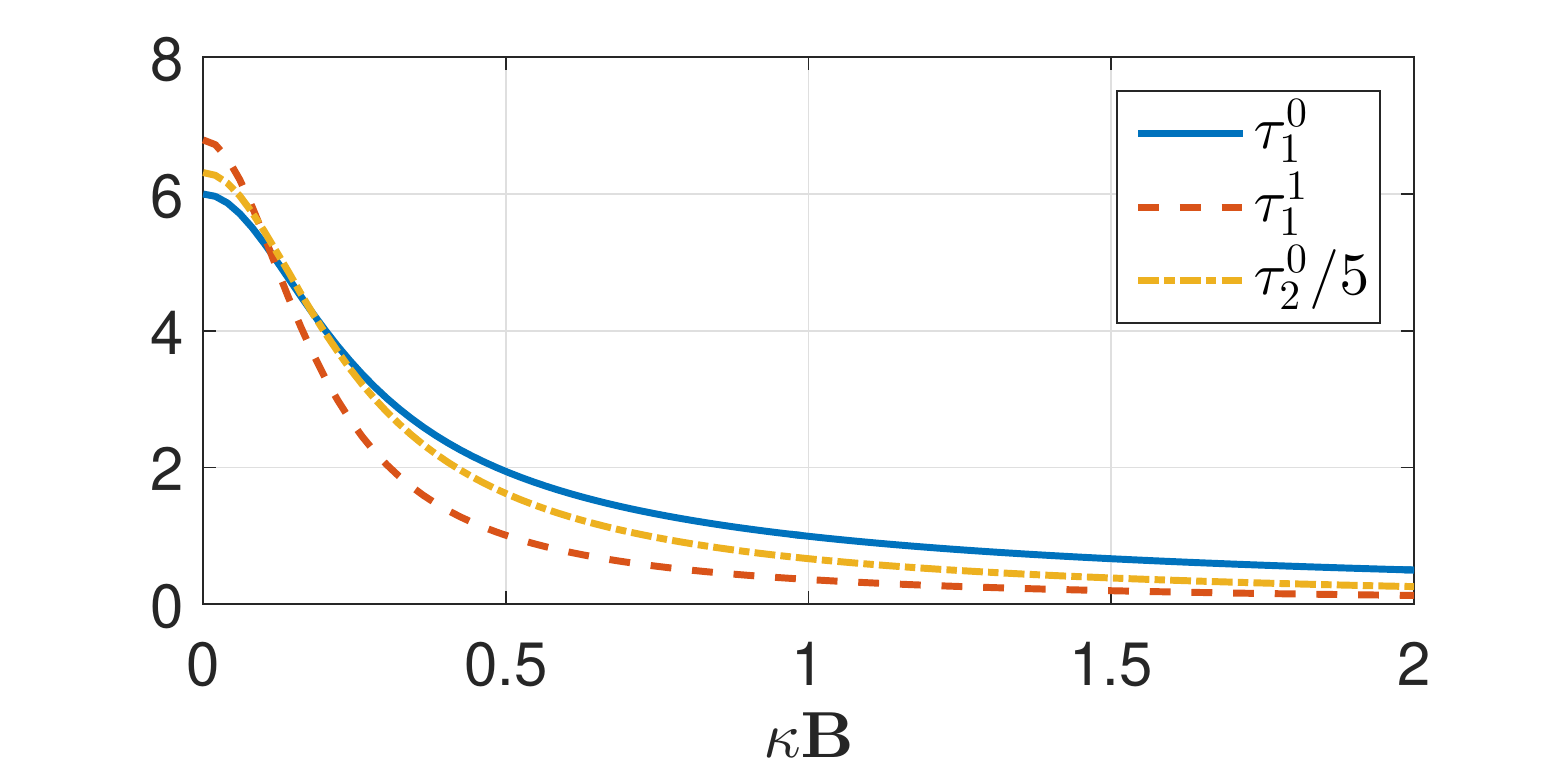}
\includegraphics[width=0.496\textwidth]{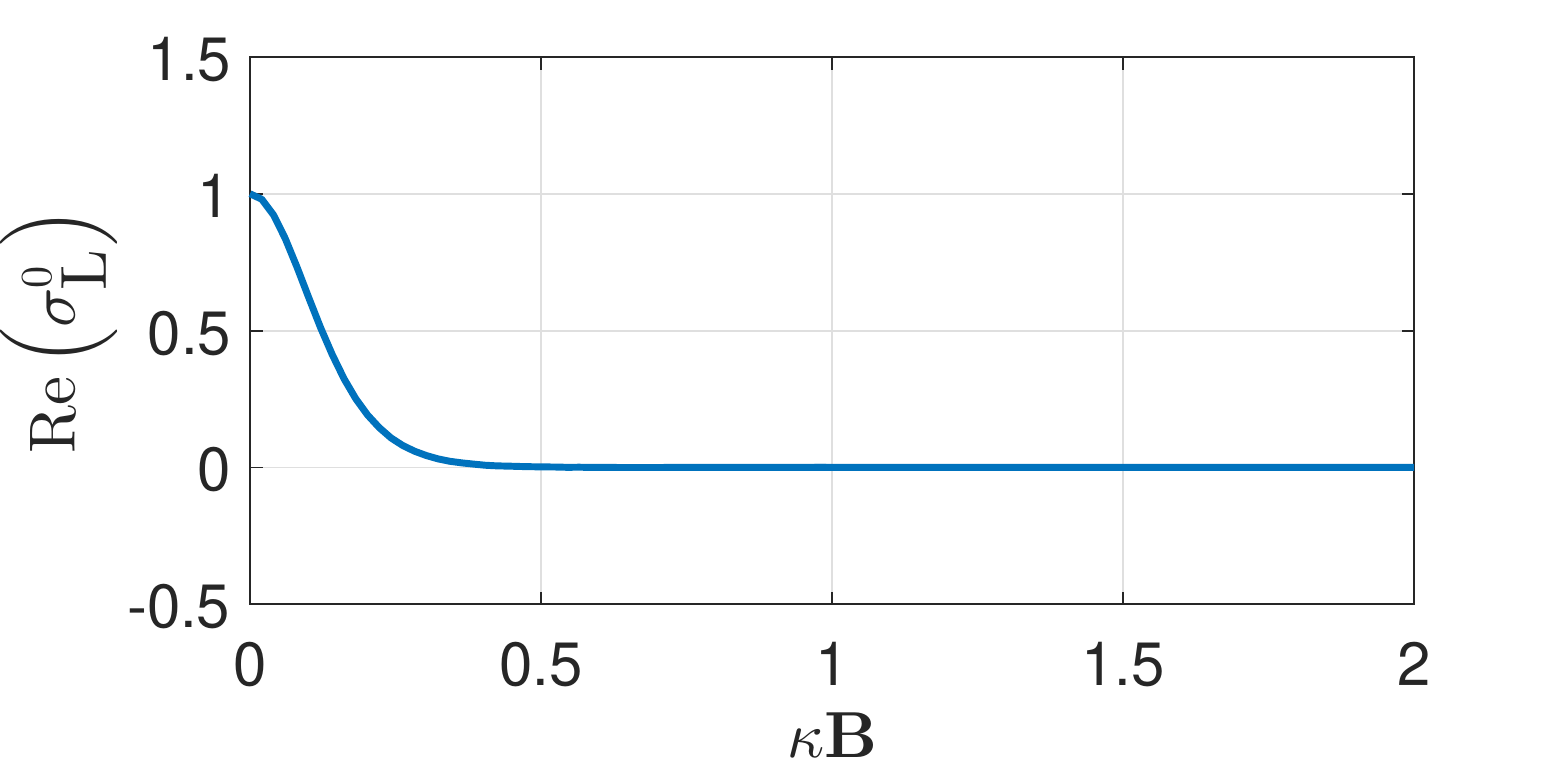}
\caption{DC conductivities $\tau_1^0$, $\tau_2^0/5$, $\tau_1^1$, and $\textrm{Re}\left(\sigma_{\textrm{L}}^0\right)$ as function of $\kappa {\bf B}$.}
\label{tauDC}
\end{figure}

To evaluate the TCFs $\tau_1,\tau_2$, we have to completely solve ODEs (\ref{eom C1}-\ref{eom C3}). We first analytically solve them when $\omega=0$. As a result, the DC limits $\tau_1^0$ (for arbitrary $\bf B$) and $\tau_2^0$ (up to leading ${\bf B}^2$-correction) are known analytically,
\begin{equation} \label{DC tau1}
\begin{split}
\tau_1^0&=\frac{\Gamma\left[3/4-\sqrt{1-144\kappa^2 {\bf B}^2}/4\right] \Gamma\left[3/4+\sqrt{1-144\kappa^2 {\bf B}^2}/4\right]}{3\kappa^2 {\bf B}^2 \Gamma\left[1/4-\sqrt{1-144\kappa^2 {\bf B}^2}/4\right] \Gamma\left[1/4+\sqrt{1-144\kappa^2 {\bf B}^2}/4\right]}\\
&\longrightarrow 6+ 216\left(1-2\log 2\right) \kappa^2 {\bf B}^2+\mathcal{O}({\bf B}^4),~~~\textrm{as}~~~{\bf B}\to 0,\\
\end{split}
\end{equation}
\begin{equation} \label{DC tau2}
\tau_2^0=18\left(\pi-2\log 2\right)+\#_2 \kappa^2 {\bf B}^2 +\mathcal{O}\left({\bf B}^4\right),~~~\textrm{as}~~{\bf B}\to 0,
\end{equation}
where $\Gamma[z]$ is a Gamma function, and $\#_2$ is known numerically only
\begin{equation}
\#_2\equiv \int_1^\infty \frac{dr}{r^3} \left\{\int_r^\infty \frac{72^2xdx}{x^4-1} \int_1^x \frac{dy}{y}\left[\log \frac{(1+y)^2}{1+y^2}-2\arctan(y)+\pi\right]\right\}
\approx-495.268.
\end{equation}
When the magnetic field is very strong, $\tau_1^0$ and $\tau_2^0$ behave similarly
\begin{equation} \label{DC strongB}
\tau_1^0,~\tau_2^0 \longrightarrow \frac{1}{\kappa {\bf B}},~~~~\textrm{as}~~\kappa {\bf B} \to \infty.
\end{equation}
The result for $\tau_1^0$ is in agreement with \cite{1410.6399,1607.06817}. In the DC limit $\omega \to 0$, when the magnetic field is very strong the on-shell vector current (\ref{jmu os}) behaves as
\begin{equation}
J^i\to -12\kappa {\bf B} \mathcal{V}_i,
\end{equation}
which is in agreement with \cite{1607.06817}. When $\omega\to 0$ (DC limit), the current-current correlator is dominated by the chiral anomaly induced effects $\sim \tau_1^0$. The DC limit is of interest for experiments with electric fields turned on adiabatically, such as the ones considered in \cite{1607.06817}. Meanwhile, when $\omega\to 0$, longitudinal conductivity $\sigma_{\textrm{L}}$ in (\ref{jmu os}) is parametrised as
\begin{equation} \label{sigma_L0}
\sigma_{\textrm{L}}^0=\frac{i}{\omega} 12\kappa^2 {\bf B}^2\tau_1^0+\left[\sigma_e^0+ \kappa^2 {\bf B}^2 \left(\tau_2^0-12\tau_1^1\right)\right],
\end{equation}
where $\sigma_e^0=1$, $\tau_1^1$ is the coefficient of $i\omega$ in hydrodynamic expansion of $\tau_1$. For illustration, in Figure \ref{tauDC} we show $\kappa {\bf B}$-dependence of $\tau_1^0$, $\tau_2^0$ (divided by 5 to match scales),
$\tau_1^1$ and $\textrm{Re}\left(\sigma_{\textrm{L}}^0\right)$. The behaviour of
$\textrm{Re}\left(\sigma_{\textrm{L}}^0\right)$ agrees perfectly with that of \cite{1410.6399}.

In our calculation, $\textrm{Re}\left(\sigma_{\textrm{L}}^0\right)$ acquires negative correction due to magnetic field and eventually vanishes when the magnetic field gets large, see Figure \ref{tauDC}. This is in contrast with many related studies of negative magnetoresistivity, the phenomenon of enhancement of longitudinal DC conductivity due to  magnetic field \cite{Nielsen:1983rb,1206.1627,1307.6990,1312.0027,Huang:2015eia,Li:2016}. However, taking strict DC limit in $\sigma_{\textrm{L}}^0$ is problematic due to the explicit $1/\omega$ divergence. The latter is frequently regularised by introduction of axial charge dissipation effects via shifting the frequency $\omega\to\omega+i/\tau_5$, where $\tau_5$ corresponds to some relaxation time. The physics of this axial charge relaxation is beyond the scope of the present work. It was addressed within the holographic approach in \cite{1407.8162,1410.6399,1504.06566,1603.02624}.
These studies primarily rely on the Kubo formula.

\begin{figure}[htbp]
\centering
\includegraphics[width=0.496\textwidth]{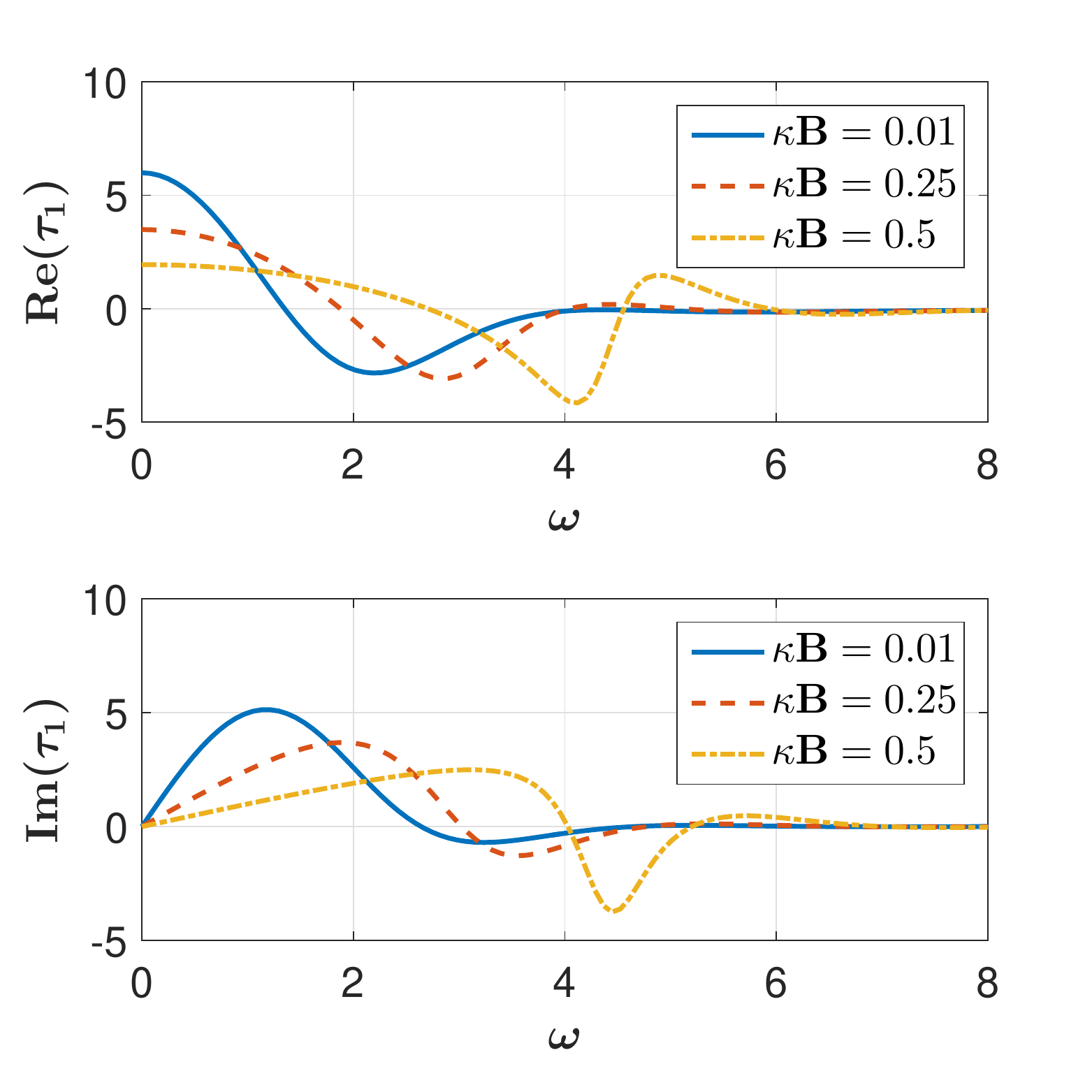}
\includegraphics[width=0.496\textwidth]{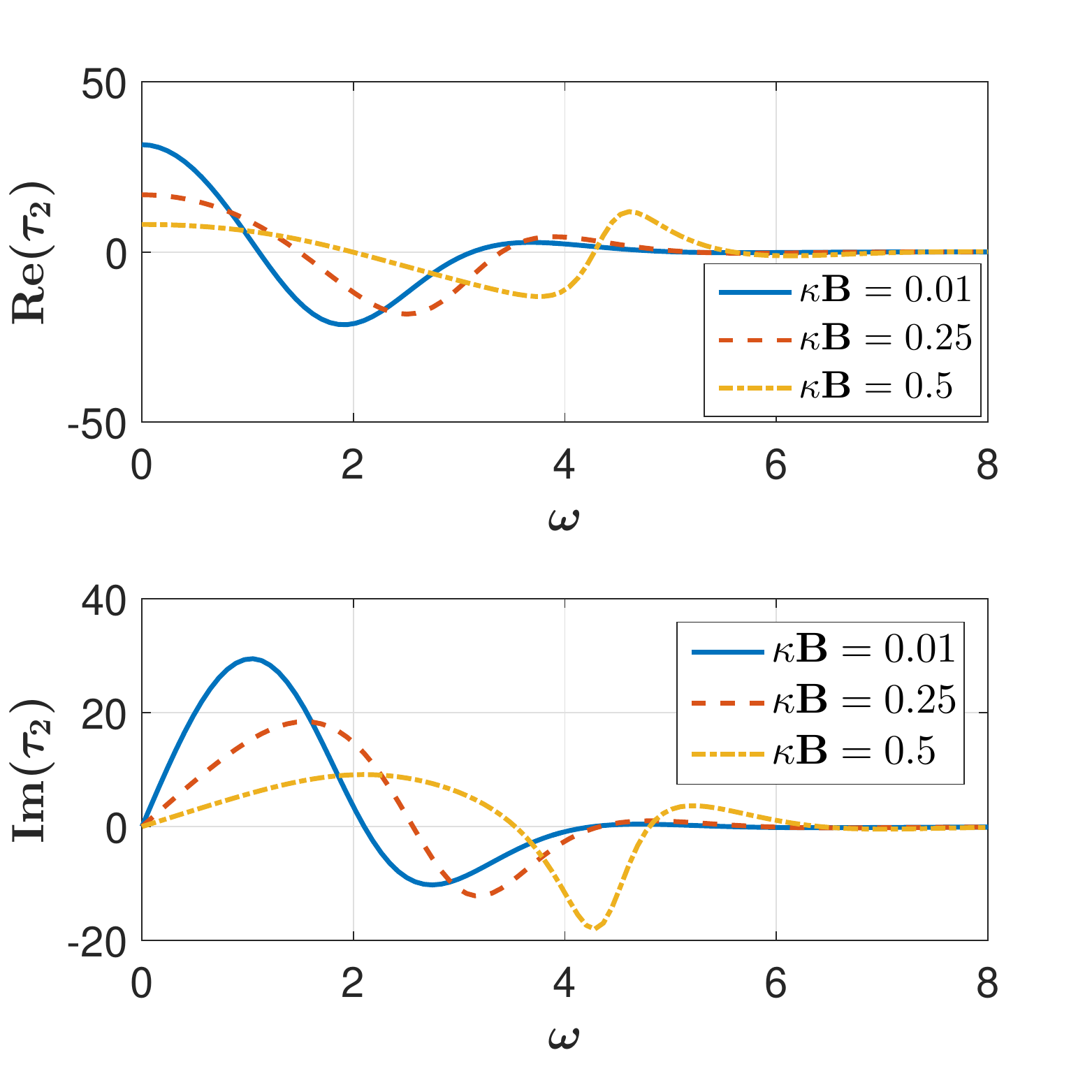}
\caption{AC conductivities $\tau_1$ and $\tau_2$ for different values of $\kappa {\bf B}$.}
\label{tau12}
\end{figure}

\begin{figure}[htbp]
\centering
\includegraphics[width=0.496\textwidth]{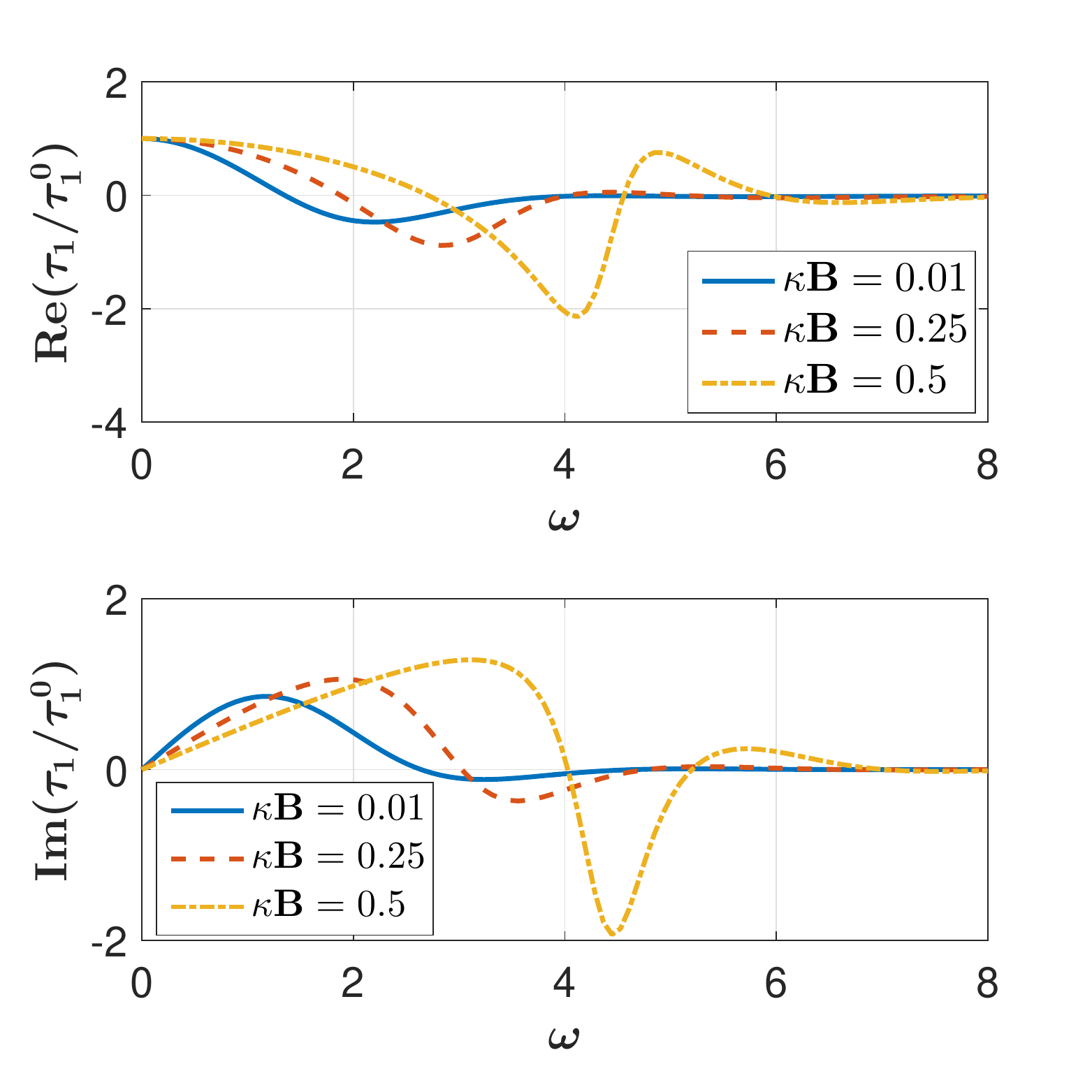}
\includegraphics[width=0.496\textwidth]{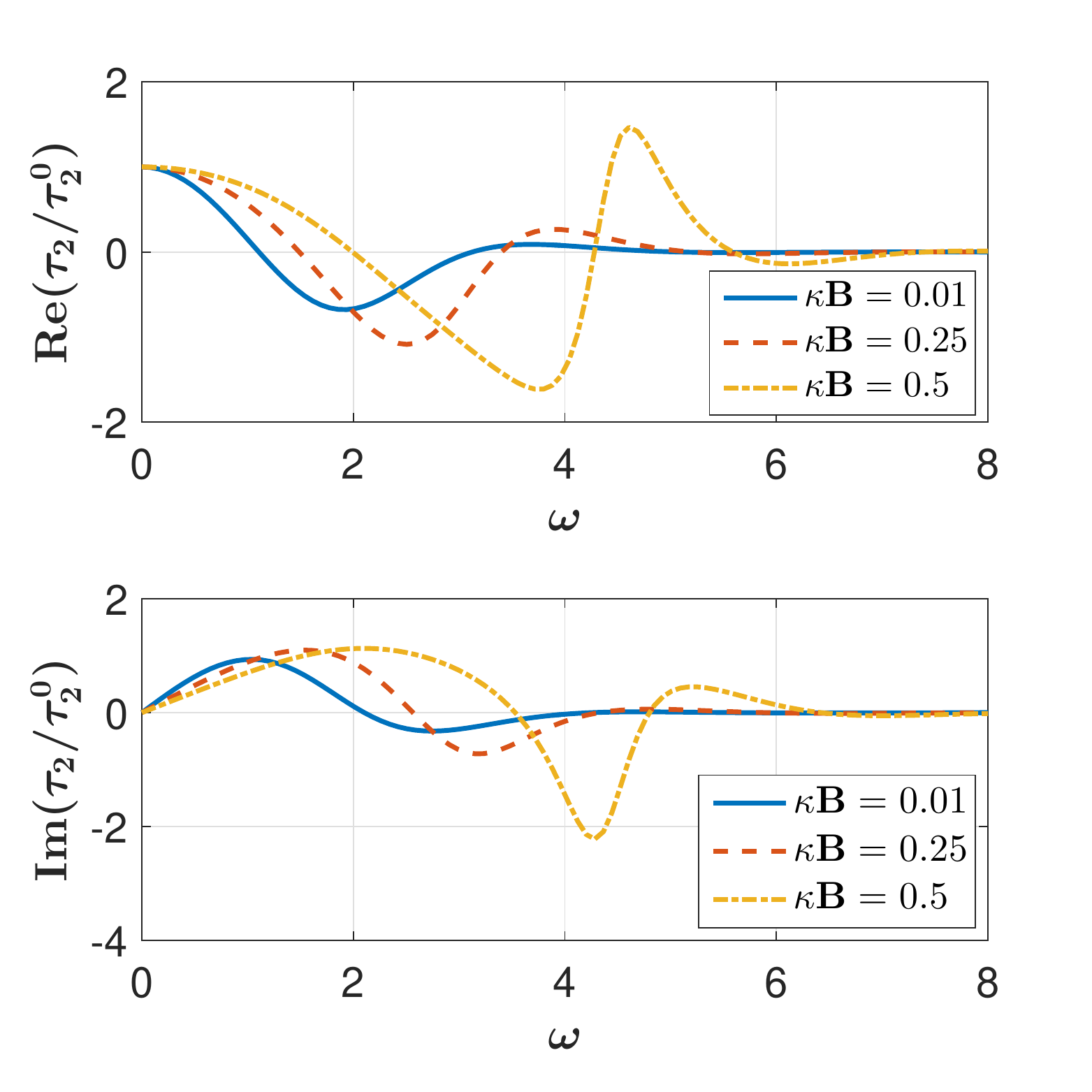}
\caption{Normalised AC conductivities $\tau_1/\tau_1^0$ and $\tau_2/\tau_2^0$ for different values of $\kappa {\bf B}$.}
\label{tau12 ratio}
\end{figure}

For arbitrary $\omega$, we resort to numerical methods and solve ODEs (\ref{eom C1}-\ref{eom C3}) for representative values of $\kappa {\bf B}$. The numerical procedure is identical to that of \cite{1608.08595} and for all the numerical details we refer the reader to this publication. In Figure \ref{tau12} we show $\omega$-dependence for $\tau_1$ and $\tau_2$ for sample choices of $\kappa {\bf B}$. In Figure \ref{tau12 ratio} we plot the normalised TCFs $\tau_1/\tau_1^0$ and $\tau_2/\tau_2^0$. Overall, $\tau_1$ and $\tau_2$ display quite similar dependence on the frequency $\omega$. After some
oscillations, both $\tau_1$ and $\tau_2$ approach zero asymptotically.

Approach to the asymptotic regime, however, depends on strength of the magnetic field. When $\kappa {\bf B}$ is increased, the asymptotic behaviour is delayed towards larger $\omega$. What is  more intriguing is that increasing $\kappa {\bf B}$ renders $\tau_1$ and $\tau_2$ to develop a resonance-like enhancement at finite $\omega$. This could be an interesting experimentally observable feature. For very strong magnetic fields $\kappa {\bf B}\to\infty$, the chiral anomaly-induced effects would be pushed to the UV,  corresponding to  early time effects, such as in  \cite{1607.06817}.

\begin{figure}[htbp]
\centering
\includegraphics[width=0.496\textwidth]{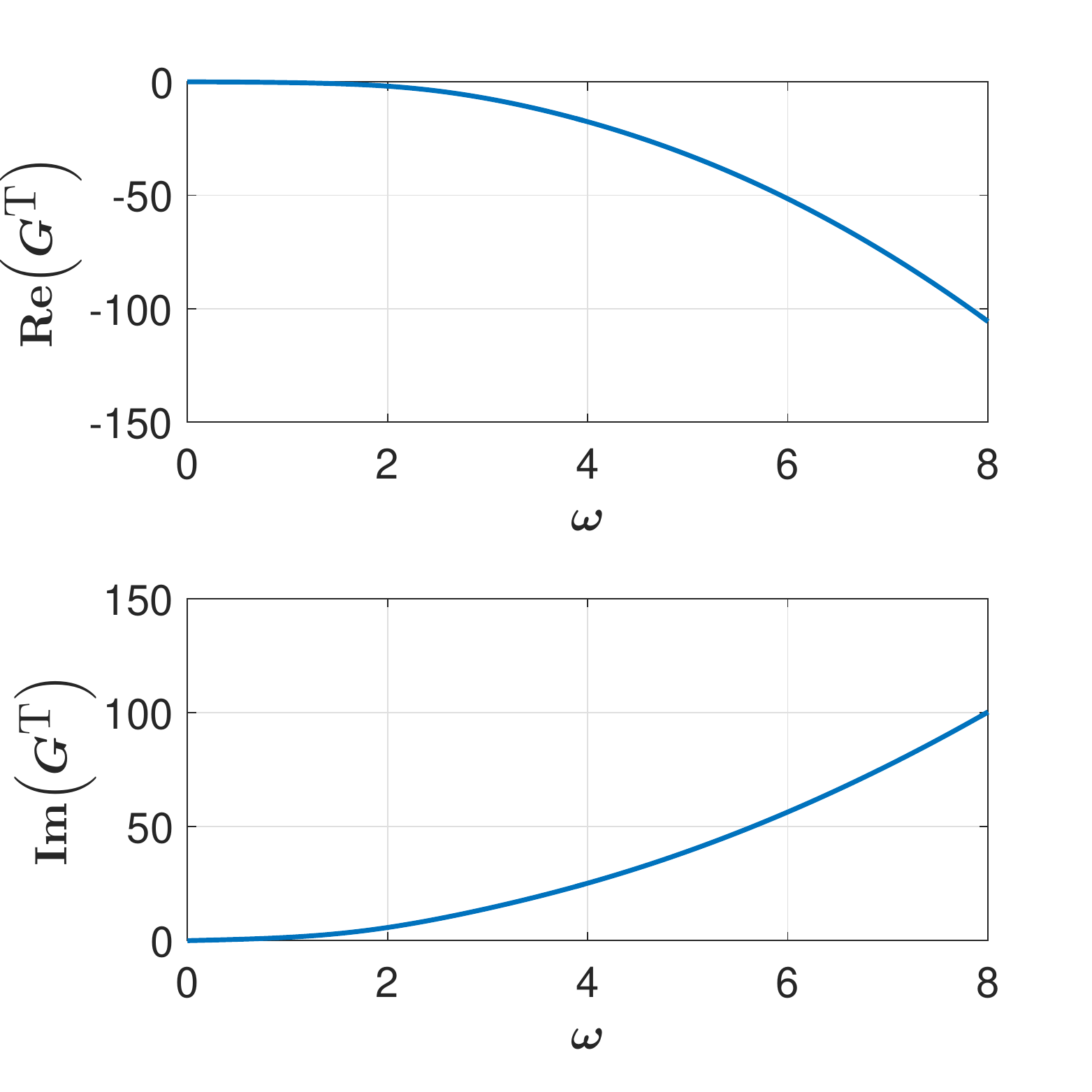}
\includegraphics[width=0.496\textwidth]{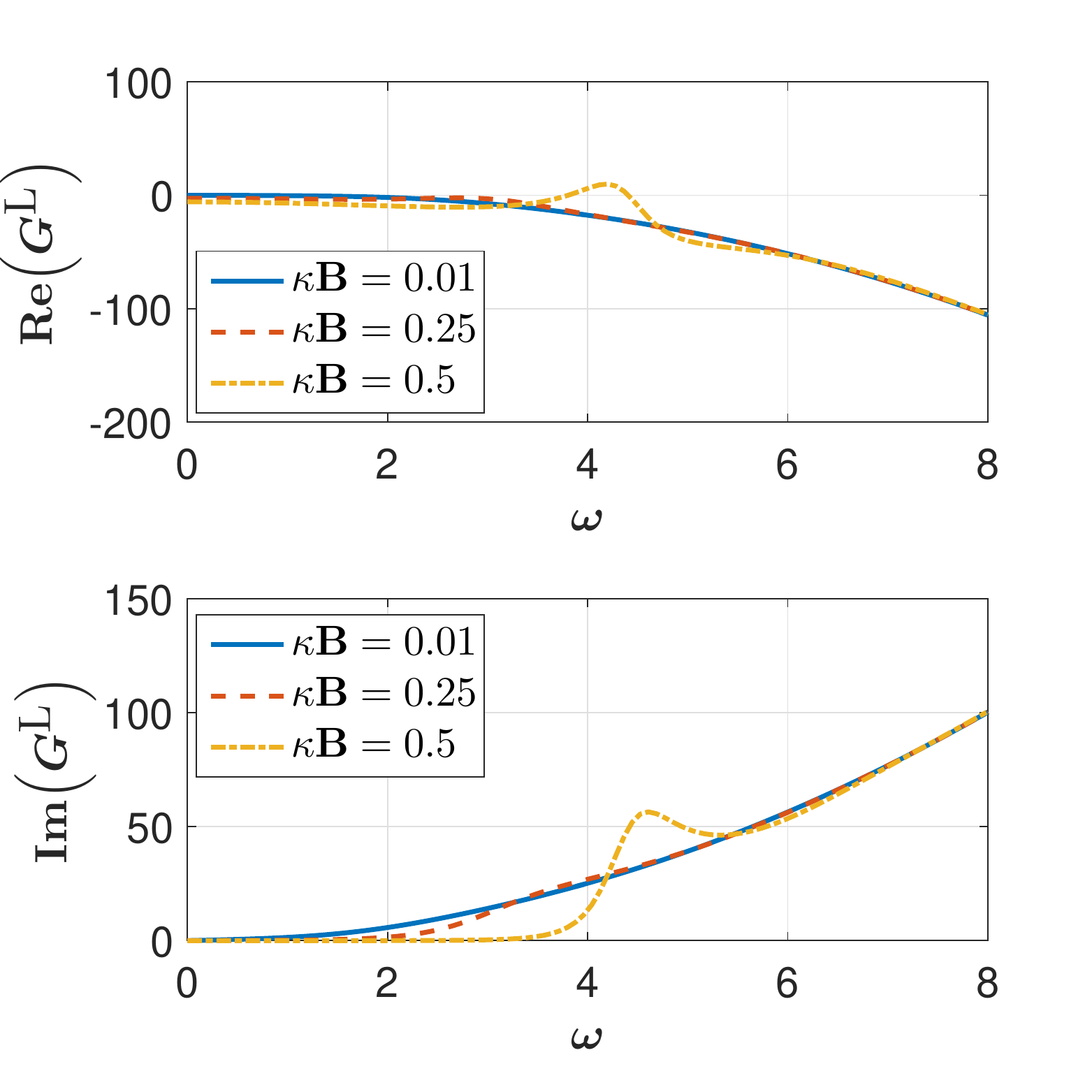}
\caption{Current-current correlators $G^{\textrm{T}}$ (left) and $G^{\textrm{L}}$ (right).}
\label{gTL}
\end{figure}

\begin{figure}[htbp]
\centering
\includegraphics[width=0.496\textwidth]{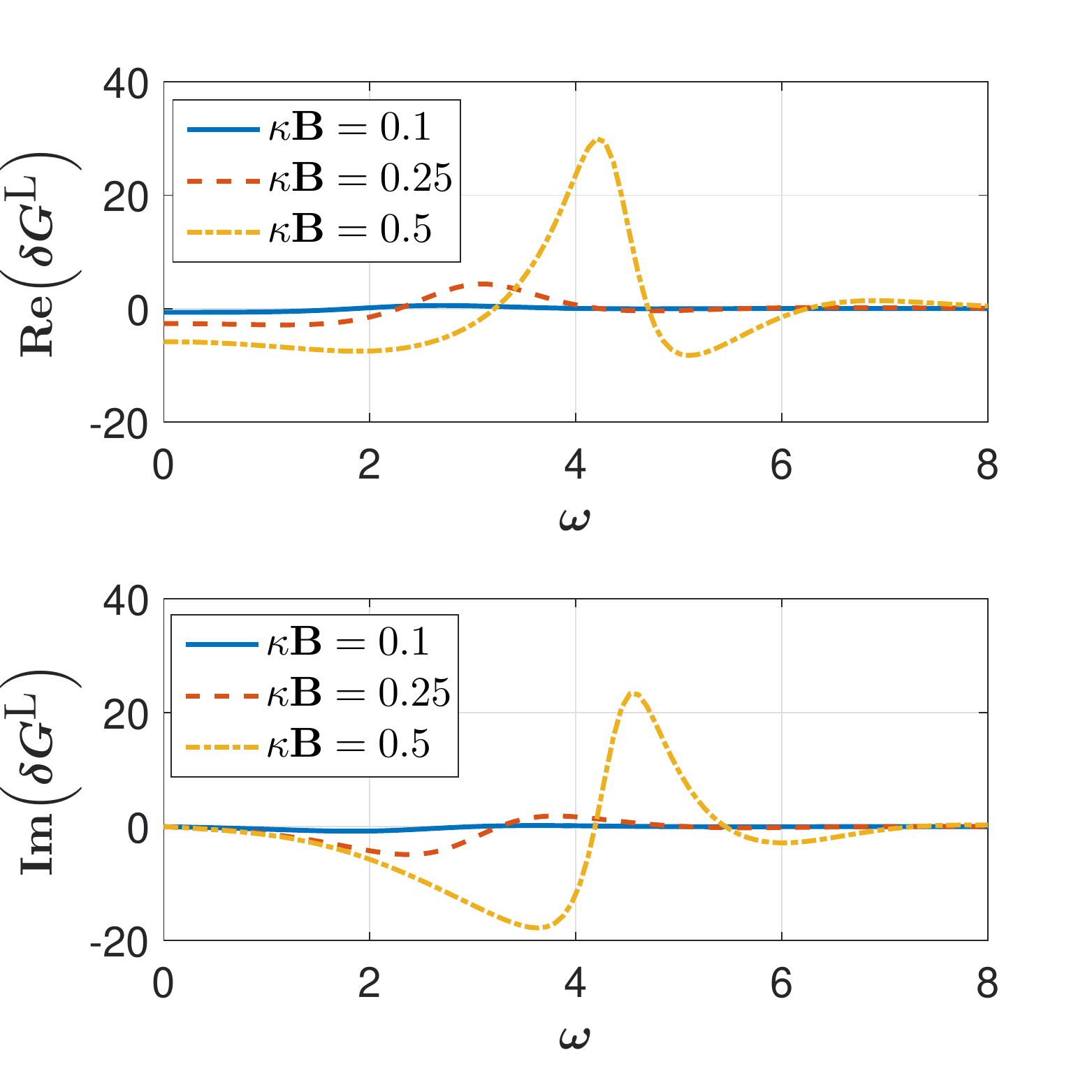}
\includegraphics[width=0.496\textwidth]{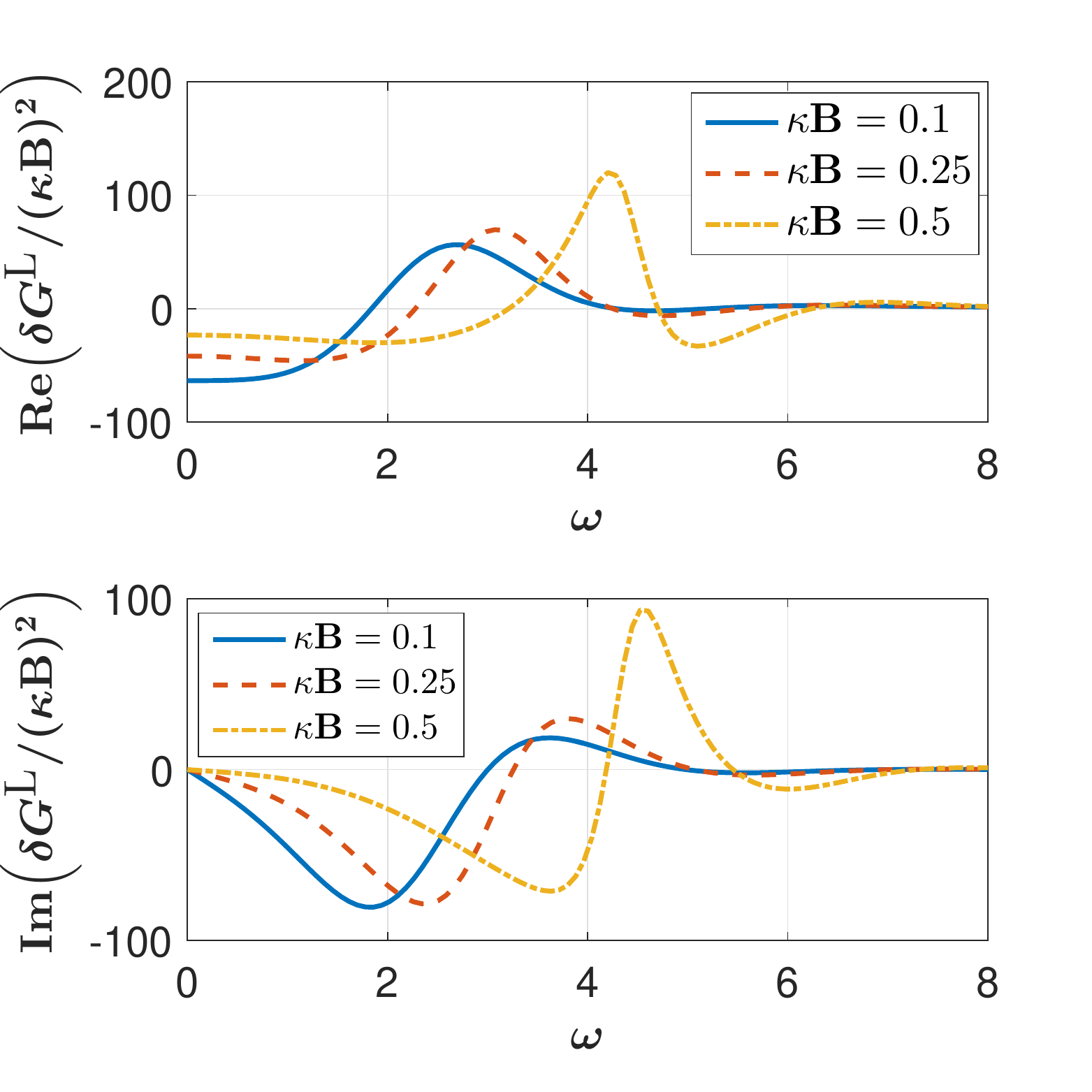}
\caption{Anomalous correction to correlator: $\delta G^{\textrm{L}}$ (left) and $\delta G^{\textrm{L}}/(\kappa {\bf B})^2$ (right).}
\label{deltagL}
\end{figure}

In Figure \ref{gTL} we show two-point correlators $G^{\textrm{T,L}}$ for different choices of $\kappa {\bf B}$. However, it is difficult to appreciate the anomaly induced effects from Figure \ref{gTL}  because in the correlators they get mixed with non--anomalous ones. To illuminate $\kappa {\bf B}$-correction to $G^{\textrm{L}}$, in Figure \ref{deltagL} we plot the difference
$\delta G^{\textrm{L}}=G^{\textrm{L}}-G^{\textrm{T}}$. From these plots, the effect of chiral anomaly on the induced vector current is seen more clearly.  We again notice a
remarkable relative enhancement  at intermediate values of $\omega$.

\section{Conclusions} \label{conclusion}

In this paper we continued  explorations of the chiral anomaly induced transport within a holographic model containing two $U(1)$ fields interacting via Chern-Simons terms. For a finite temperature system, we computed off-shell constitutive relations for the vector/axial currents responding to external electromagnetic fields.

When a static spatially-inhomogeneous magnetic field is the only external field that is turned on, we showed that both the  CME and CSE get corrected by derivative terms, see (\ref{jmu varying B1},\ref{jmu5 varying B1}). Within the derivative expansion, we analytically calculated corrections up to third order in the expansion, see (\ref{Gi cme1},\ref{Hi cme1},\ref{Gi 3rd},\ref{Hi 3rd}). Apart from the derivative corrections to CME and CSE,  the diffusion constant $\mathcal{D}_0$ was found to receive a negative anomaly-induced correction, see (\ref{anomalous D0}). The dispersion relation of the chiral magnetic wave was also found to be modified, see(\ref{dispersion}).

In the second part of our study, we focused on the case of time-varying electric and constant magnetic fields without any externally enforced axial charge asymmetry,
though the $\vec{E}(t)\cdot \vec{\bf B}$ term in the continuity equation (\ref{continuity1}) generates the axial charge density $\rho_{_5}$ (and thus $\mu_{_5}$) dynamically. For such configuration of external fields, we first analysed the most general constitutive relations for the vector/axial currents, see (\ref{jmu exp},\ref{jmu5 exp}). Then, within the derivative expansion, we explicitly calculated the currents up to third order at nonlinear level, see (\ref{jmu exp gradient},\ref{jmu5 exp gradient}). When put on-shell, the axial current $\vec{J}_5$ is fully nonlinear in the external electric field.

Employing another approximation, we linearised the constitutive relations assuming  the  electric field is weak (\ref{amplitude exp}). Within this approximation the axial current is zero, while the ``off-shell'' vector current is parameterised by three frequency-dependent transport coefficient functions: the electric conductivity $\sigma_e$, and two chiral anomaly-induced conductivities $\tau_1,\tau_2$, see (\ref{current exp}). In the DC limit, we analytically computed these conductivities, see (\ref{DC tau1},\ref{DC tau2}). Then, for generic $\omega$, the numerical plots were presented in section \ref{AC conductivities}. Based on these studies, we notice that the anomaly-induced effects get enhanced at some finite frequency $\omega$, whereas the position of the maximum and strength of the effect depends on the external magnetic field.  It might be an effect worth looking for experimentally.

\appendix

\section{Supplement for  Section \ref{cme}} \label{app cme}

In this Appendix we provide computational details regarding non-renormalisation of CME and its gradient corrections up to third order. The dynamical equations (\ref{eom Vt}-\ref{eom Ai}) have a special property: in all equations, the first two terms  can be rewritten as total derivatives of $\mathbb{V}_\mu,\mathbb{A}_\mu$. Treating all the remaining terms in (\ref{eom Vt}-\ref{eom Ai}) as sources,  (\ref{eom Vt}-\ref{eom Ai}) can be integrated over $r$ twice, resulting in the following integral forms
\begin{equation} \label{Vt cme}
\begin{split}
\mathbb{V}_t(r)&=-\int_r^\infty \frac{dx}{x^3} \int_x^\infty \left\{y \partial_y\partial_k \mathbb{V}_k +12 \kappa \epsilon^{ijk}\left[ \partial_y \mathbb{A}_i \left(\partial_j \mathcal{V}_k +\partial_j \mathbb{V}_k \right)+ \partial_y \mathbb{V}_i \partial_j \mathbb{A}_k\right]\right\}dy,\\
&\underrightarrow{r\to \infty}\, \mathcal{O}\left(\frac{\log r}{r^3}\right),
\end{split}
\end{equation}
\begin{align} \label{Vi cme}
\mathbb{V}_i(r)&=\int_r^\infty \frac{-xdx}{x^4-1} \left\{G_i(x)- \frac{1-x}{2x} \partial_i\rho - \partial_k \mathcal{F}_{ki}^V\log x - 12\kappa B_i \left(\mu_{_5}+ \mathbb{A}_t-\frac{1}{2x^2} \rho_{_5} \right) \right\}\nonumber\\
&\underrightarrow{r\to\infty}\, -\frac{\partial_i \rho}{4r^2} +\frac{1+2\log r}{4r^2} \partial_k\mathcal{F}_{ki}^V+ \frac{6}{r^2} \kappa \mu_{_5} B_i -\frac{1}{2r^2} G_i(x=\infty)+ \mathcal{O}\left(\frac{\log r}{r^3}\right),
\end{align}
\begin{equation} \label{At cme}
\begin{split}
\mathbb{A}_t(r)&=-\int_r^\infty \frac{dx}{x^3} \int_x^\infty \left\{y \partial_y \partial_k \mathbb{A}_k +12 \kappa \epsilon^{ijk}\left[ \partial_y \mathbb{V}_i \left(\partial_j \mathcal{V}_k +\partial_j \mathbb{V}_k \right)+ \partial_y \mathbb{A}_i \partial_j \mathbb{A}_k\right]\right\}dy\\
&\underrightarrow{r\to \infty}\, \mathcal{O}\left(\frac{\log r}{r^3}\right),
\end{split}
\end{equation}
\begin{equation} \label{Ai cme}
\begin{split}
\mathbb{A}_i(r)&=-\int_r^\infty \frac{xdx}{x^4-1} \left\{H_i(x)- \frac{1-x}{2x} \partial_i \rho_{_5} -12\kappa \mu B_i-12\kappa B_i \left(\mathbb{V}_t-\frac{1}{2x^2} \rho\right) \right\}\\
&\underrightarrow{r\to \infty}\,-\frac{\partial_i \rho_{_5}}{4r^2} +\frac{6}{r^2}\kappa \mu B_i-\frac{1}{2r^2}H_i(x=\infty)+\mathcal{O}\left(\frac{1}{r^3}\right),
\end{split}
\end{equation}
where $\mu$ and $\mu_{_5}$ are the chemical potentials defined in (\ref{def potentials}). The frame convention~(\ref{Landau frame}) was utilised to fix integration constants, one for $\mathbb{V}_t$ and one for $\mathbb{A}_t$. The functions $G_i(x)$ and $H_i(x)$ are
\begin{equation} \label{Gi cme}
\begin{split}
G_i(x)=\int_1^x dy &\left\{-2y\partial_y \partial_t \mathbb{V}_i +y\partial_y \partial_i \mathbb{V}_t-\left(\partial_t \mathbb{V}_i- \partial_i \mathbb{V}_t\right)- \frac{1}{y} \left(\partial^2\mathbb{V}_i -\partial_i \partial_k \mathbb{V}_k\right)\right.\\
&\left.-12\kappa \epsilon^{ijk} \partial_y\left(\mathbb{A}_t-\frac{1}{2y^2} \rho_{_5} \right) \partial_j \mathbb{V}_k-12\kappa \epsilon^{ijk} \partial_y \left(\mathbb{V}_t -\frac{1}{2y^2} \rho \right) \partial_j \mathbb{A}_k\right.\\
&\left.+12\kappa \epsilon^{ijk} \partial_y \mathbb{A}_j\left[\left( \partial_t \mathbb{V}_k- \partial_k \mathbb{V}_t\right) +\frac{1}{2y^2} \partial_k \rho\right]+12\kappa \epsilon^{ijk} \partial_y \mathbb{V}_j\right.\\
&\left.\times \left[\left( \partial_t \mathbb{A}_k- \partial_k \mathbb{A}_t\right) +\frac{1}{2y^2} \partial_k \rho_{_5}\right] \right\},
\end{split}
\end{equation}
\begin{equation} \label{Hi cme}
\begin{split}
H_i(x)=\int_1^x dy &\left\{-2y\partial_y \partial_t \mathbb{A}_i +y\partial_y \partial_i \mathbb{A}_t-\left(\partial_t \mathbb{A}_i- \partial_i \mathbb{A}_t\right)- \frac{1}{y} \left(\partial^2\mathbb{A}_i -\partial_i \partial_k \mathbb{A}_k\right)\right.\\
&\left.-12\kappa \epsilon^{ijk} \partial_y\left(\mathbb{V}_t-\frac{1}{2y^2} \rho \right) \partial_j \mathbb{V}_k-12\kappa \epsilon^{ijk} \partial_y \left(\mathbb{A}_t -\frac{1}{2y^2} \rho_{_5} \right) \partial_j \mathbb{A}_k\right.\\
&\left.+12\kappa \epsilon^{ijk} \partial_y \mathbb{V}_j\left[\left( \partial_t \mathbb{V}_k- \partial_k \mathbb{V}_t\right) +\frac{1}{2y^2} \partial_k \rho\right]+12\kappa \epsilon^{ijk} \partial_y \mathbb{A}_j\right.\\
&\left.\times\left[\left( \partial_t \mathbb{A}_k- \partial_k \mathbb{A}_t\right) +\frac{1}{2y^2} \partial_k \rho_{_5}\right] \right\}.
\end{split}
\end{equation}
Although we were unable to solve (\ref{eom Vt cme1}-\ref{eom Ai cme1}) for generic $\rho, \rho_{_5}, \vec{B}(\vec{x})$, integral forms (\ref{Vt cme}-\ref{Ai cme}) help to explore general forms of $J^\mu/J_5^\mu$, as quoted in (\ref{jmu varying B1},\ref{jmu5 varying B1}).

In the hydrodynamic limit, we analytically solved the dynamical equations (\ref{eom Vt cme1}-\ref{eom Ai cme1}) within the boundary derivative expansion (\ref{derivative exp}). $\mathbb{V}_\mu^{[n]}$ and $\mathbb{A}_\mu^{[n]}$ up to $n=2$ are listed below.
\begin{equation} \label{pert VAt 1st}
\mathbb{V}_t^{[1]}=\mathbb{A}_t^{[1]}=0,
\end{equation}
\begin{equation} \label{pert Vi 1st}
\mathbb{V}_i^{[1]}=-\frac{1}{8}\left[\log \frac{1+r^2}{(1+r)^2}-2\arctan(r)+\pi\right] \partial_i \rho + 3\kappa \rho_{_5} B_i \log \frac{1+r^2}{r^2},
\end{equation}
\begin{equation} \label{pert Ai 1st}
\mathbb{A}_i^{[1]}=\mathbb{V}_i^{[1]}\left(\rho\leftrightarrow \rho_{_5}\right),
\end{equation}
\begin{equation} \label{pert Vt 2nd}
\begin{split}
\mathbb{V}_t^{[2]}&=-\int_r^\infty \frac{dx}{x^3} \int_x^\infty dy \left\{ \frac{y\,\partial^2\rho}{2(y^2+1)(y+1)}- \frac{y\, 6\kappa B_k \partial_k\rho_{_5}} {(y^2+1)(y+1)}- \frac{72\kappa^2}{y(y^2+1)}\rho B^2\right\}\\
&\xlongequal{r=1}-\frac{1}{16}\left(\pi-2\log 2\right)\partial^2\rho+ \frac{3}{4} \left(\pi-2\log 2\right) \kappa B_k\partial_k \rho_{_5}-18(1-2\log 2)\kappa^2 \rho B^2,
\end{split}
\end{equation}
\begin{equation} \label{pert At 2nd}
\mathbb{A}_t^{[2]}=\mathbb{V}_t^{[2]}\left(\rho\leftrightarrow \rho_{_5}\right),
\end{equation}
\begin{equation} \label{pert Vi 2nd}
\begin{split}
\mathbb{V}_i^{[2]}&=b_0 \partial_k \mathcal{F}_{ki}^V+ b_1 \partial_t\partial_i \rho+ b_2 6\kappa \partial_t \rho_{_5} B_i+ b_3 36\kappa^2 \epsilon^{ijk} \left[\rho_{_5} \partial_j\left(\rho_{_5} B_k\right)+ \rho \partial_j\left(\rho B_k\right)\right]\\
&+b_4 36\kappa^2 \epsilon^{ijk} \left(\rho B_j \partial_k \rho+ \rho_{_5} B_j \partial_k \rho_{_5}\right),
\end{split}
\end{equation}
\begin{equation} \label{pert Ai 2nd}
\begin{split}
\mathbb{A}_i^{[2]}&=b_1 \partial_t \partial_i \rho_{_5}+ b_2 6\kappa \partial_t \rho B_i+ b_3 36\kappa^2 \epsilon^{ijk} \left[\rho\partial_j\left(\rho_{_5}B_k\right)+ \rho_{_5} \partial_j\left(\rho B_k\right)\right]+b_4 36\kappa^2 \epsilon^{ijk} \\
&\times \left(\rho_{_5} B_j \partial_k \rho+ \rho B_j \partial_k \rho_{_5}\right),
\end{split}
\end{equation}
where
\begin{equation}
b_0=\int_r^\infty \frac{xdx}{x^4-1} \int_1^x \frac{dy}{y},
\end{equation}
\begin{equation}
b_1=\int_r^\infty \frac{xdx}{x^4-1} \int_1^x dy \left\{\frac{y}{(y^2+1)(y+1)}- \frac{1}{8}\left[\log\frac{1+y^2}{(1+y)^2}- 2\arctan(y)+\pi\right] \right\},
\end{equation}
\begin{equation} \label{b2}
b_2=\int_r^\infty \frac{xdx}{x^4-1} \int_1^x dy \left\{-\frac{2}{y^2+1}+ \frac{1}{2} \log \frac{1+y^2}{y^2}\right\},
\end{equation}
\begin{equation}
b_3=\int_r^\infty \frac{xdx}{x^4-1} \int_1^x \frac{1}{y^3}\log \frac{1+y^2}{y^2} dy,
\end{equation}
\begin{equation}
b_4=\int_r^\infty \frac{xdx}{x^4-1} \int_1^x \frac{dy}{y^3(y^2+1)}.
\end{equation}
These perturbative solutions, once inserted into (\ref{Gi cme}, \ref{Hi cme}, \ref{def potentials}), produces the results (\ref{Gi cme1}-\ref{Hi 3rd}).

\section{Supplement for Section \ref{exp setup}} \label{app exp}

This Appendix contains calculational details for section \ref{exp setup}.  As explained in Appendix \ref{app cme}, integrating the dynamical equations (\ref{eom Vt expf}-\ref{eom Ai expf}) over $r$ twice results in the following integral forms,
\begin{equation} \label{Vt exp}
\mathbb{V}_t=12\kappa \int_r^\infty \frac{dx}{x^3} \mathbb{A}_k {\bf B}_k\,\underrightarrow{r\to \infty}\, \mathcal{O}\left(\frac{1}{r^3}\right),
\end{equation}
\begin{equation} \label{Vi exp}
\begin{split}
\mathbb{V}_i&=\int_r^\infty \frac{xdx}{x^4-1}\int_1^x dy \left\{2y\partial_y \partial_t \mathbb{V}_i +\partial_t \mathbb{V}_i -E_i+ 12\kappa \partial_y \left(\mathbb{A}_t -\frac{1}{2y^2} \rho_{_5}\right){\bf B}_i\right.\\
&~~~~~~~~~~~~~~~~~~~~~~~~~~~~~~-\left.12\kappa \epsilon^{ijk} \partial_y \mathbb{A}_j \left(-E_k+ \partial_t \mathbb{V}_k \right)-12\kappa \epsilon^{ijk} \partial_y \mathbb{V}_j \partial_t \mathbb{A}_k\right\}\\
&\underrightarrow{r\to \infty}\,\frac{1+2\log r}{4r^2}\partial_tE_i-\left(\frac{1}{r}- \frac{1}{2r^2}\right)E_i+\frac{6}{r^2}\kappa \mu_{_5} {\bf B}_i -\frac{6}{r^2}\kappa \epsilon^{ijk} \mathbb{A}_j(1)E_k\\
&~~~~~~~~~+\frac{1}{2r^2} \overline{G}_i(x=\infty)+\mathcal{O}\left(\frac{1}{r^3} \right),
\end{split}
\end{equation}
\begin{equation} \label{At exp}
\mathbb{A}_t=12\kappa \int_r^\infty \frac{dx}{x^3} \mathbb{V}_k {\bf B}_k\,\underrightarrow{r\to \infty}\, \mathcal{O}\left(\frac{1}{r^3}\right),
\end{equation}
\begin{equation} \label{Ai exp}
\begin{split}
\mathbb{A}_i&=\int_r^\infty \frac{xdx}{x^4-1}\int_1^x dy \left\{2y\partial_y \partial_t \mathbb{A}_i +\partial_t \mathbb{A}_i + 12\kappa \partial_y \mathbb{V}_t {\bf B}_i+12\kappa \epsilon^{ijk} \partial_y \mathbb{V}_j E_k\right.\\
&~~~~~~~~~~~~~~~~~~~~~~~~~~~~~~\left.-12\kappa \epsilon^{ijk} \partial_y \mathbb{V}_j \partial_t \mathbb{V}_k -12\kappa \epsilon^{ijk} \partial_y \mathbb{A}_j \partial_t \mathbb{A}_k\right\}\\
&\underrightarrow{r\to \infty}\,\frac{6}{r^2}\kappa \mu {\bf B}_i -\frac{6}{r^2}\kappa \epsilon^{ijk} \mathbb{V}_j(1)E_k+\frac{1}{2r^2} \overline{H}_i(x=\infty)+\mathcal{O} \left(\frac{1}{r^3}\right),
\end{split}
\end{equation}
where $\mu,\mu_{_5}$ are defined via (\ref{def potentials}). $\overline{G}_i$ and $\overline{H}_i$ are
\begin{equation} \label{Gi bar}
\overline{G}_i(x)=\int_1^x dy \left\{2y\partial_y \partial_t \overline{\mathbb{V}}_i+ \partial_t \overline{\mathbb{V}}_i -12\kappa \epsilon^{ijk} \left(\partial_y \mathbb{A}_j \partial_t \mathbb{V}_k + \partial_y \mathbb{V}_j \partial_t \mathbb{A}_k \right) \right\},
\end{equation}
\begin{equation} \label{Hi bar}
\overline{H}_i(x)=\int_1^x dy \left\{2y\partial_y \partial_t \mathbb{A}_i+ \partial_t \mathbb{A}_i -12\kappa \epsilon^{ijk} \left(\partial_y \mathbb{V}_j \partial_t \mathbb{V}_k + \partial_y \mathbb{A}_j \partial_t \mathbb{A}_k \right) \right\},
\end{equation}
where $\overline{\mathbb{V}}_i=\mathbb{V}_i+E_i/r$. Note that we have split the $E_i/r$ piece from $\mathbb{V}_i$ so that $\overline{G}_i(x)$ is well defined at $x=\infty$. Via the general formulas (\ref{bdry currents}), the large $r$ behaviors (\ref{Vt exp}-\ref{Ai exp}) produce the formal results of (\ref{jmu exp},{\ref{jmu5 exp}}).

Under the boundary derivative expansion (\ref{derivative exp}), (\ref{eom Vt exp}-\ref{eom Ai exp}) can be solved perturbatively. Up to second order $\mathcal{O}(\partial^2)$, the corrections $\mathbb{V}_\mu$ and $\mathbb{A}_\mu$ are
\begin{equation} \label{Vt cme 2nd}
\mathbb{V}_t=\mathcal{O}\left(\partial^3\right),
\end{equation}
\begin{equation} \label{At cme 2nd}
\mathbb{A}_t=a_0(r)12\kappa \vec{E}\cdot \vec{\bf B}- \frac{18}{r^2}\left[1-(1+r^2)\log \frac{1+r^2}{r^2}\right]\kappa^2 \rho_{_5} {\bf B}^2+\mathcal{O}\left(\partial^3\right),
\end{equation}
\begin{equation} \label{Vi cme 2nd}
\begin{split}
\mathbb{V}_i&=-\frac{1}{4}\left[\log \frac{(1+r)^2}{1+r^2}-2\arctan(r)+\pi\right]E_i+ 3 \log \frac{1+r^2}{r^2} \kappa \rho_{_5}{\bf B}_i+a_1(r) \partial_t E_i\\
&+b_2(r) 6\kappa \partial_t\rho_{_5} {\bf B}_i+\mathcal{O}\left(\partial^3\right),
\end{split}
\end{equation}
\begin{equation} \label{Ai cme 2nd}
\mathbb{A}_i=a_2(r)72\kappa^2 \rho_{_5}\epsilon^{ijk} {\bf B}_jE_k + \mathcal{O} \left(\partial^3\right),
\end{equation}
where
\begin{equation}
a_0(r)=-\frac{1}{8r^2}\left\{(r^2+1)\left(2\,\mathrm{arccot}(r)- \log\frac{1+r^2}{r^2} \right) +2(r^2-1) \log \frac{r}{1+r} \right\},
\end{equation}
\begin{equation}
a_1(r)=-\int_r^\infty\frac{xdx}{x^4-1}\int_1^x dy\left\{-\frac{2y^2}{1+y^2} +\frac{1}{4}\left[\log \frac{(1+y)^2}{1+y^2} -2\arctan(y)+\pi\right]\right\},
\end{equation}
\begin{equation}
a_2(r)=-\int_r^\infty\frac{xdx}{x^4-1} \int_1^x \frac{dy}{y(y^2+1)},
\end{equation}
and $b_2$ is presented in (\ref{b2}). From these perturbative results, we deduce the hydro expansion for the currents $J^\mu/J_5^\mu$ and chemical potentials $\mu/\mu_{_5}$, as summarised in (\ref{jmu exp gradient},\ref{jmu5 exp gradient},\ref{mu/mu5 exp}).

\section*{Acknowledgements}
We would like to thank Dmitri E. Kharzeev, Alex Kovner, Andrey Sadofyev, Derek Teaney, and Ho-Ung Yee for useful discussions related to this work. YB would like to thank KITPC (Beijing) for financial support and hospitality, Physics Department of the University of Connecticut for hospitality where part of this work was done. This work was supported by the ISRAELI SCIENCE FOUNDATION grant \#1635/16, BSF grant \#012124, the People Program (Marie Curie Actions) of the European Union's Seventh Framework under REA grant agreement \#318921; and the Council for Higher Education of Israel under the PBC Program of Fellowships for Outstanding Post-doctoral Researchers from China and India (2015-2016).

\providecommand{\href}[2]{#2}\begingroup\raggedright\endgroup

\end{document}